\documentclass[twocolumn]{aastex62}

\usepackage{amsmath}
\usepackage{graphicx}
\usepackage{natbib}
\usepackage{txfonts}
\usepackage{multirow}
\usepackage{bm}
\usepackage{booktabs}
\usepackage{lineno}
\shorttitle{ 21 cm BAO }
\shortauthors{ Zhang \& Chan et al. }

\newcommand{\Ddel}{\delta_{\rm D}   }

\newcommand{\MpcOh}{ \,  \mathrm{Mpc} \,  h^{-1} }
\newcommand{\hOMpc}{ \,  \mathrm{Mpc}^{-1}   h  }

\newcommand{\comment}[1]{}
\newcommand{\nn}{ \nonumber }

\newcommand{\beq}{\begin{equation}}
\newcommand{\eeq}{\end{equation}}
\newcommand{\beqa}{\begin{eqnarray}}
\newcommand{\eeqa}{\end{eqnarray}}

\newcommand{\HI}{${\rm H}_{\rm I}$ }

\newcommand{\change}[1]{\textcolor{black}{#1}}

\begin{document}

%\title{ A forecast on the 21 cm BAO measurement with different statistics }
\title{ Recovery of 21 cm Baryonic Acoustic Oscillations: A Configuration-space Correlation Function Analysis }
%\title{ Normalizing flow approach to photo-z estimation  }

\correspondingauthor{Kwan Chuen Chan}
\email{chankc@mail.sysu.edu.cn}

\author{Zhongyue Zhang}
\affiliation{ School of Physics and Astronomy, Sun Yat-Sen University, 2 Daxue Road, Tangjia, Zhuhai 519082, China }
\affiliation{ CSST Science Center for the Guangdong-Hongkong-Macau Greater Bay Area, SYSU, Zhuhai 519082, China }

\author{Kwan Chuen Chan}
\affiliation{ School of Physics and Astronomy, Sun Yat-Sen University, 2 Daxue Road, Tangjia, Zhuhai 519082, China }
\affiliation{ CSST Science Center for the Guangdong-Hongkong-Macau Greater Bay Area, SYSU, Zhuhai 519082, China }

\author{Santiago Avila}
\affiliation{Centro de Investigaciones Energ\'eticas, Medioambientales y Tecnol\'ogicas (CIEMAT), 28040 Madrid, Spain}

\author{Bernhard Vos-Gin\'es}
\affiliation{Fakult\"at f\"ur Physik, Universit\"at Bielefeld, Postfach 100131, 33501 Bielefeld, Germany}

%% \correspondingauthor{Kwan Chuen Chan}
%% \email{chankc@mail.sysu.edu.cn}

%% \author{Kwan Chuen Chan}
%% \affiliation{ School of Physics and Astronomy, Sun Yat-Sen University, Guangzhou 510275, China }

\begin{abstract}
  Intensity mapping (IM) represents an innovative and potent probe to cosmology. One of its prime applications is to measure the Baryonic Acoustic Oscillations (BAO) in the late universe. We study the BAO measurement by IM in configuration space using simulations, focusing on the impact of the telescope beam and foreground removal effects, \change{which are modeled by a Gaussian beam and radial power suppression, respectively. }   Three types of correlation functions are applied to measure BAO, including the radial correlation function, multipole correlation function, and  wedge correlation function.  We check our pipeline against a set of IM mock catalogs, finding good agreement with the numerical results.  We use the mock catalogs to look for the parameter choices that optimize the BAO constraint for the correlation function estimators.  With the optimal settings, our pipeline is utilized to forecast the BAO constraint for the 21 cm IM experiments: BINGO, MeerKAT, and SKA-mid.  \change{ Despite our simple beam and foreground removal effect treatments, our analyses could offer a useful guide to strengthen the constraining power of these experiments.  }   We find that for the low-redshift experiments BINGO and MeerKAT, the wedge correlation function achieves the tightest constraint for both the transverse and radial BAO.  For SKA-mid, the radial correlation function and multipole correlation function deliver the tightest constraint for the radial and transverse BAO, respectively.  

\end{abstract}

%\keywords{editorials, notices ---
%  miscellaneous --- catalogs --- surveys}


\section{ Introduction}

21 cm intensity mapping (IM) represents a transformative, statistical approach to observational cosmology, offering a powerful probe across multiple cosmic epochs \citep{Battye_etal2013,Bharadwaj_etal2001,McQuinn_etal2006,Mao_etal2008,PritchardLoeb_2008}.  The 21 cm radiation arises from the hyperfine spin-flip transition of neutral hydrogen (${\rm H}_{\rm I}$), transitioning from the triplet electron-proton spin state to the singlet ground state.  Unlike traditional galaxy surveys, which resolve and catalog individual galaxies, IM measures the aggregate 21 cm radio emission from \HI across vast angular patches on the sky. It does so by creating a low-resolution but efficient three-dimensional map of the total \HI luminosity.

During the cosmic dawn and the subsequent epoch of reionization, IM can reveal the birth and growth of the first large-scale structures. By mapping the faint 21 cm signal from these early times, it can track how the first stars and galaxies ionize their surrounding neutral gas, illuminating one of the last major phase transitions in cosmic history \citep{ Furlanetto_etal2006,PritchardLoeb_2012}.   At lower redshifts, the neutral hydrogen probed by IM predominantly resides within galaxies. Consequently, the large-scale fluctuations in the 21 cm signal serve as an excellent tracer of the underlying cosmic web of dark matter in the post-reionization universe. This makes late-time IM a complementary and potent tool for studying contemporary large-scale structure. It provides a unique avenue to measure cosmological parameters, test models of gravity, and investigate the nature of dark energy through Baryon Acoustic Oscillations (BAO) and redshift-space distortions \citep{Change_etal_2008, LoebWyithe_2008,Bull_etal2015,Santos_etal2015, Obuljen_etal2018}. Steady progress has been made in the 21 cm signal detection. The signals have not only been measured via cross correlation with the galaxy survey tracers, e.g.,~\citet{Chang_etal2010,Masui_etal2013,Anderson_etal2018,Wolz_etal2022,Cunnington_etal2023,Amiri_etal2024}, but direct auto correlation measurements have also been claimed \citep{Paul_etal2023,Amiri_etal_CHIME2025}.

In this work, we focus on the measurement of BAO using IM. BAO are primordial acoustic features imprinted on the large-scale distribution of matter in the universe \citep{PeeblesYu1970,SunyaevZeldovich1970}. Their origin lies in the pre-recombination era, when photons and baryons (electrons and protons) are tightly coupled in a hot, dense plasma. Within this medium, pressure waves propagate until the epoch of recombination, when atoms form and the photons decouple. This process freezes the waves, leaving a characteristic signature in the spatial arrangement of matter. The scale of this signature corresponds to the sound horizon at the drag epoch—the maximum distance these sound waves could travel, approximately 150 Mpc in standard cosmology. Since the physics governing BAO formation is linear and well-modeled within the early universe, this sound horizon scale can be calculated with high precision. Consequently, BAO serves as a robust standard ruler for cosmology, enabling precise measurements of the universe's expansion history and geometry (e.g.~\cite{BondEfstathiou1984,BondEfstathiou1987,HuSugiyama1996,HuSugiyamaSilk1997,Dodelson_2003}).
Ever since its clear detection in SDSS \citep{Eisenstein_etal2005}  and 2dFGS \citep{Cole_etal2005}, BAO measurements have become a routine and essential component of large-scale structure analyses. They have been precisely charted across cosmic time by a succession of major large-scale structure survey analyses, providing a growing and powerful probe for constraining cosmological models \citep{Gaztanaga:2008xz, Beutler_etal2011, Anderson_BOSS2012, Kazin_etal2014, Alam_etal2017, DES_Y1BAO2019, eBOSS:2020yzd, DES_Y3BAO2022,Chan_DESY3BAO2022, DES_Y6BAO2024, DESI_BAO1_2025,DESI_BAO2_2025}.

Radio surveys for cosmology employ two primary measurement strategies. The first utilizes arrays of antennae functioning as an interferometer, whose angular resolution is determined by the minimum and maximum separation between its elements. However, many planned interferometric surveys are not optimized for BAO measurements. For instance, SKA will not be packed tight enough to offer high sensitivity for BAO \citep{Santos_etal2015}.  In contrast, the single-dish mode, where each antenna independently observes a large patch of sky, is well-suited for BAO science. This approach naturally provides the wide, contiguous field of view and high mapping speed needed to efficiently capture the vast scales of the BAO signal \citep{Battye_etal2013}. For this reason, key ongoing and upcoming 21 cm IM experiments, including BINGO \citep{Abdalla_etal2022}, MeerKAT \citep{Santos_etal2016}, and the SKA Observatory \citep{Bacon_SKA2020}, are strategically designed to utilize this method, offering prime opportunities to harness BAO as a cosmological probe. The IM BAO is probably not as competitive as the measurement from galaxy surveys, but it paves the way for further IM studies in cosmology, e.g.,~the measurement of primordial non-Gaussianity.

However, there are two observational challenges, the telescope beam effect and the foreground removal effect, that can contaminate the signals. The telescope beam causes the modes to be smoothed in the transverse direction. Because  the 21 cm signal is a few orders of magnitude smaller than the strong astrophysical foreground, aggressive cleaning is required, and this also removes the long-wavelength modes in the radial direction.     This work studies the impact of these systematic effects on the recovery of the BAO signals via the correlation function approach. \change{ Because of the simplicity of the treatment of the anisotropy effects in Fourier space, most 21 cm IM BAO studies focus on Fourier space, e.g.,~\citet{Bull_etal2015,Villaescusa-Navarro_etal2017,Li_etal2025}.  However, the correlation function analysis can serve as important cross-check and some of the effects, such as the survey mask can be handled more easily in configuration space. So far, there are only limited studies in configuration space \citep{KennedyBull_2021,Avila_etal2022,Novaes_etal2022,Bizarria_etal}. }   \citet{KennedyBull_2021} applied the multipole correlation function to investigate the 21 cm IM BAO measurement. In \citet{Avila_etal2022}, a set of 21 cm IM simulation mocks was built and various configuration space correlation functions were measured, but the BAO fitting was not performed. Here, we systematically study the recovery of the BAO through three different correlation function estimators. We first check our pipeline against the simulations of \citet{Avila_etal2022}. After verifying our pipeline, we go on to forecast the BAO constraints for various 21 cm IM experiments.

This paper is organized as follows. In Sec.~\ref{sec:correlation_functions}, we review the correlation function formalism and present three types of correlation function estimators for 21 cm IM studies. Sec.~\ref{sec:covariances} is devoted to the covariance of the estimators. In Sec.~\ref{sec:method}, we introduce the simulation sets for verification of the pipeline and then describe the method used to infer the parameters.  Sec.~\ref{sec:results} presents the main results. In this section, we first use the mock catalog to look for the optimal parameter choices yielding the most robust and constraining results  under various beam width and foreground removal strength.  We go on to use the optimal setting to forecast the BAO constraint for the major ongoing and forthcoming 21 cm IM experiments. We conclude in Sec.~\ref{sec:conclusions}. Unless otherwise stated, the cosmology adopted is a flat $\Lambda$CDM with cosmological parameters $\Omega_{\rm m}=0.309, \, h=0.667, \, n_{\rm s}=0.967, \, \sigma_8 =0.815$ \citep{Planck2016}.

\section{ 21 cm correlation function statistics }
\label{sec:correlation_functions}

In this section, we first introduce the model for 21 cm IM power spectrum and correlation function, and then describe three types of correlation functions used to measure the 21 cm IM clustering signal in this work. 
    
\subsection{21 cm power spectrum}
\label{sec:21cm_Pk}

%\KCC{ Check the standard Pk splitting  }

We use  {\tt camb} \citep{Lewis_etal2000} to compute the underlying nonlinear matter power spectrum implemented with the halo model fit \citep{Smith_etal2003}. The particular halo model fit adopted is \citet{Mead_etal2021}, which, among other things, includes accurate BAO damping.
 BAO damping is the smearing of the BAO feature due to the large-scale velocity dispersion, and it can be approximated by a Gaussian damping factor. In \citet{Mead_etal2021}, the damping factor is computed using the perturbation theory result from \citet{CrocceScoccimarro2006}. We have checked that the results are in agreement with those computed by other means, such as by \citet{Ivanov_etal2018}. \change{  We note that the damping can be partially undone in the 21 cm IM case using the technique of reconstruction \citep{Obuljen_etal2017}.  }

In terms of the matter power spectrum $P_{\rm m}$, the cosmological 21 cm fluctuation power spectrum is given by
\begin{equation}
  P_{\rm 21 cm }^{\rm cosmo}(k,z)=b^{2}_{\rm 21 cm}(z)P_{\rm m}(k,z),
\end{equation}
where $b_{\rm 21 cm}$ denotes $\bar{T}_{\rm b}  b_{ {\rm H}_{\rm I} } $ with $\bar{T}_{\rm b}$ being the mean brightness temperature and  $b_{ {\rm H}_{\rm I} } $ the H$_{\rm I }$ bias parameter. The brightness temperature can be expressed as \citep{Battye_etal2013} 
\beq
\bar{T}_{\rm b} = 191  {\rm mK } \,  \frac{ H_0 (1+z)^{2}}{ H(z)} \Omega_{\rm H_{I} }(z) h,   
\eeq
where $H(z)$  is the Hubble parameter at $z$ and $H_0 = 100 \, h \, \mathrm{km \, s^{-1} \, Mpc^{-1}} $. The neutral hydrogen density parameter $ \Omega_{\rm H_{I} }(z)$ is defined to be the mean neutral hydrogen density at redshift $z$, normalized by the {\it present} critical density.

On top of the cosmological 21 cm signals, there are two important observational effects causing significant distortion to the observed signals.  First, in the case of single-dish IM experiments, the telescope beam causes smearing in the transverse direction.  We take the beam to be Gaussian, and its effect on the Fourier mode is modeled by a Gaussian damping factor
\begin{equation}
  \label{eq:Rbeam_model}
  B_{\rm beam}(k_{\bot})=\exp \Bigg(-\frac{k_{\bot}^2R^2_{\rm beam}}{2} \Bigg),
\end{equation}
where $k_\perp $ denotes the wavenumber transverse to the line of sight.  The width of the Gaussian beam $ R_{\rm beam}$ is defined as \citep{Villaescusa-Navarro_etal2017}
\begin{equation}
  R_{\rm beam}=\frac{\theta_{\rm FWHM}}{\sqrt{8 \ln2}}r(z), 
\end{equation}
where $r(z)$ is the comoving distance to the source and $\theta_{\rm FWHM}$ is the angular full width at half maximum of the beam, which is  computed as
\beq
\theta_{\rm FWHM} = \frac{ \lambda}{ D_{\rm dish} }, 
\eeq
with $\lambda$ being the redshifted 21 cm signal and $D_{\rm dish}$  the dish diameter.

The second distortion is related to the foreground contamination. The cosmological 21 cm signals suffer from serious foreground contamination caused by the galactic synchrotron radiation, free-free emission, and extragalactic sources. This is particularly challenging because the foreground is a few orders of magnitude higher than the signals. Fortunately, the foreground contamination is smooth in frequency, and this property is often exploited to clean the 21 cm IM map; see e.g.,~\citet{Wolz_etal2014, Alonso_etal2015, LiuShaw_2020, Cunnington_etal2021} for some of the methods proposed to mitigate the issue. However, the cleaning unavoidably also removes the smooth cosmological signals. We model the effect of the foreground removal on the Fourier modes by a damping term in Fourier space as 
\begin{equation}
  \label{eq:Bfg_model}
  B_{\rm fg}(k_{\parallel}) = 1- \exp \Bigg(- \frac{k^2_{\parallel}}{k^2_{\rm fg}} \Bigg) , 
\end{equation}
where $k_\parallel $ denotes the wavenumber along the line-of-sight direction and  $k_{\rm fg} $ signifies the characteristic wavenumber at which foreground removal imparts a significant effect.  We parametrize $ k_{\rm fg} $  by % \citep{Soares_etal2021} 
\begin{equation}
  \label{eq:kfg_prescription}
  k_{\rm fg}=N_{\parallel}\frac{2\pi}{L_{z}}, 
\end{equation}
where  $L_z$  is the radial length scale of the survey, and  $N_{\parallel}$ emulates the number of fundamental modes removed by the foreground removal method. \citet{Soares_etal2021} empirically find that  $N_{\parallel} = 2 $ can approximate well  $N_{\rm IC} = 4 $ independent components removed by the FastICA method \citep{Hyvarinen_1999}. In this work, we thus take $N_{\parallel} = 2 $. \change{  While Eq.~\eqref{eq:Bfg_model} captures the power suppression of the foreground removal in the radial modes and has been adopted in previous works, e.g.,~\cite{KennedyBull_2021,Avila_etal2022}, we note that it neglects complications including mode mixing, foreground wedges, and potential residuals after cleaning \citep{Morales_etal2012,Liu_etal2014,LiuShaw_2020}.  We limit ourselves to this simple model and leave it to future work to further quantify the impacts of these effects. Nevertheless, we will partially test the resilience of the model by adopting a foreground removal  model different from the underlying one in Sec.~\ref{sec:Impact_Mismodeling}. }

Hence, in the presence of beam  and foreground removal effect, the observed 21 cm IM  power spectrum model reads
\begin{equation}
  \label{eq:Pk_21cmmodel}
  P_{\rm 21 cm}(k_{\parallel},k_{\bot})=b^{2}_{\rm 21 cm}B_{\rm beam}^2(k_{\perp})B^2_{\rm fg}(k_{\parallel})P_{m}(k,z). 
\end{equation}

In the first part of this work, we check our model against the simulation of  \citet{Avila_etal2022}, which aims toward modeling an SKA-like survey, and here we follow their parameter choice. In the second part, we will use the survey specifications of BINGO, MeerKAT, and SKA to forecast their BAO constraints.
\citet{Avila_etal2022} considered two scenarios for the radial size of the survey: $L_{z}=3450 \MpcOh$ and $L_{z}=300\MpcOh$, corresponding to the full SKA range $0.3< z <3$ and a small redshift bin $1.1< z <1.3$. Following \citet{Avila_etal2022}, we use two beam choices, $R_{\rm beam}= 10$ and $38.45\MpcOh$, with the latter corresponding to the SKA beam and the former to a lower-redshift study or a bigger dish size.  In total, including the systematics-free cases, we discuss the scenarios with  $R_{\rm beam} =0, \, 10, \, 38.45  \MpcOh$ and $k_{\rm fg }= 0, \, 0.00364, \, 0.0419  \hOMpc$, respectively. As \citet{Avila_etal2022}, the mean brightness temperature  $ \bar{T}_{\rm b} $ is set to 0.183 mK. 
%the density parameter for neutral hydrogen is set to  $\Omega_{\rm H_{\rm I}} = 8.7 \times 10^{-4}$.
The \HI bias parameter  $ b_{ \rm H_{\rm I}} $ is taken to be  $1.48$, which is obtained by fitting to the isotropic correlation function from the simulation. In the second part, $\Omega_{\rm H_{\rm I} }  $ and $ b_{ \rm H_{\rm I}} $ are computed using the prescription in \citet{Villaescusa-Navarro_etal2017} (Eq.~12 in that work), which are in very good agreement with the values adopted in the first part.

\change{ The redshift space distortion (RSD) gives another source of anisotropy, and on large scales, it would contribute a factor of  $B_{\rm RSD}^2$ to  $ P_{\rm 21 cm} $ in Eq.~\eqref{eq:Pk_21cmmodel}:
\beq
B_{\rm RSD} ( k_{\parallel},k_{\bot}) =  1 + \beta \mu^2,   
\eeq
where $ \beta = f/ b_{\rm H_{ I} } $ with  $f $ being the linear growth rate.  In the first part of this work,  we do not consider the RSD effect, as the simulations from \citet{Avila_etal2022} were performed in real space to concentrate on the anisotropies specific to 21 cm IM, but we will include it in the forecast part.  }

\begin{figure*}
  \includegraphics[width=0.96\linewidth]{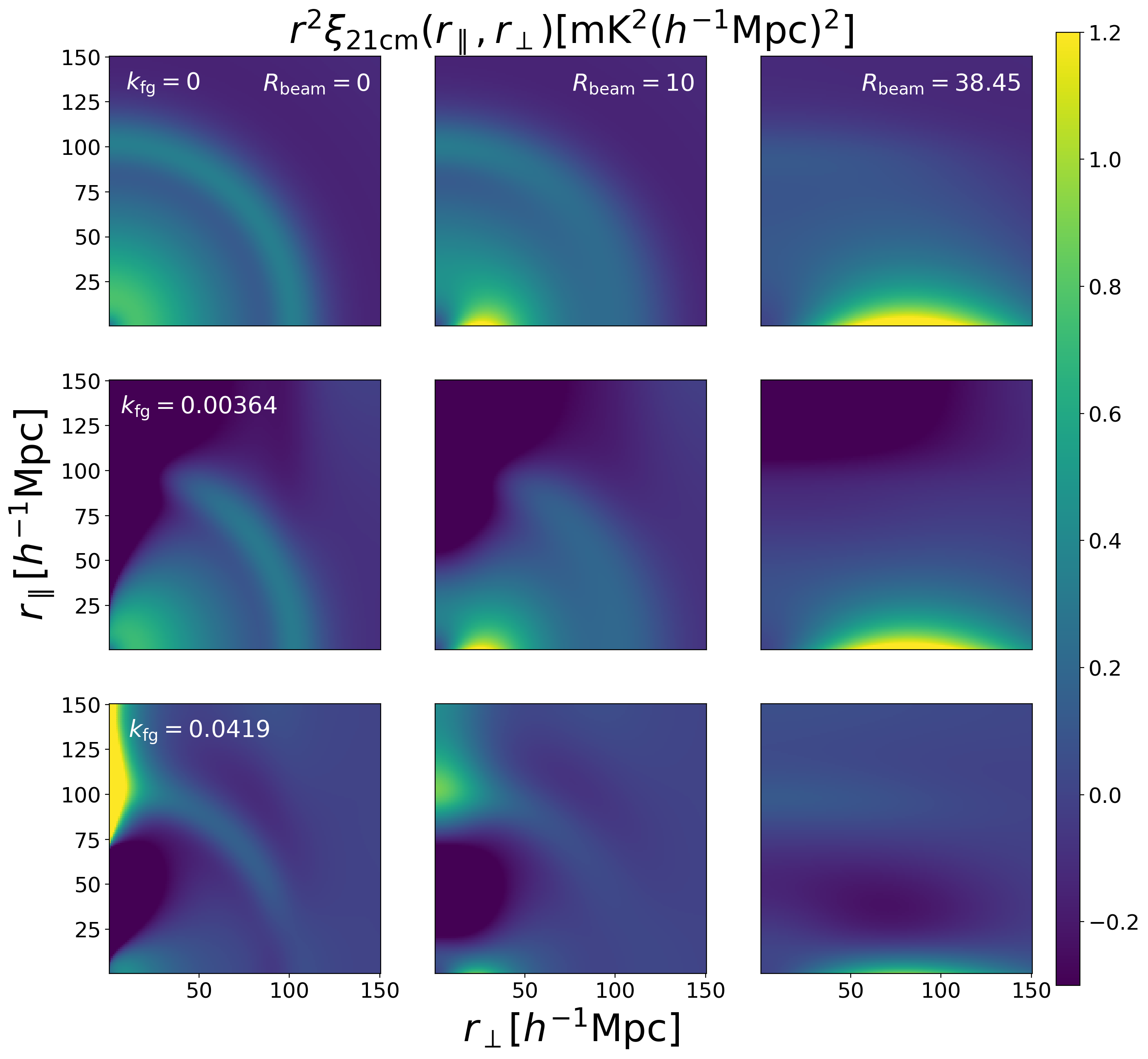}
\caption{ 2D correlation function $\xi_{\rm 21cm} $ computed with Eq.~\eqref{eq:xi21cm} for different values of $R_{\rm beam}$ (0, 10 and 38.45 $\MpcOh$ from left to right) and $k_{\rm fg}$ (0, 0.00364, 0.0419 $\hOMpc$ from top to bottom). See the text for more details. }
    \label{fig:xi2D_model}
\end{figure*}

\begin{figure*}[!htb]
  \includegraphics[width=0.98\linewidth]{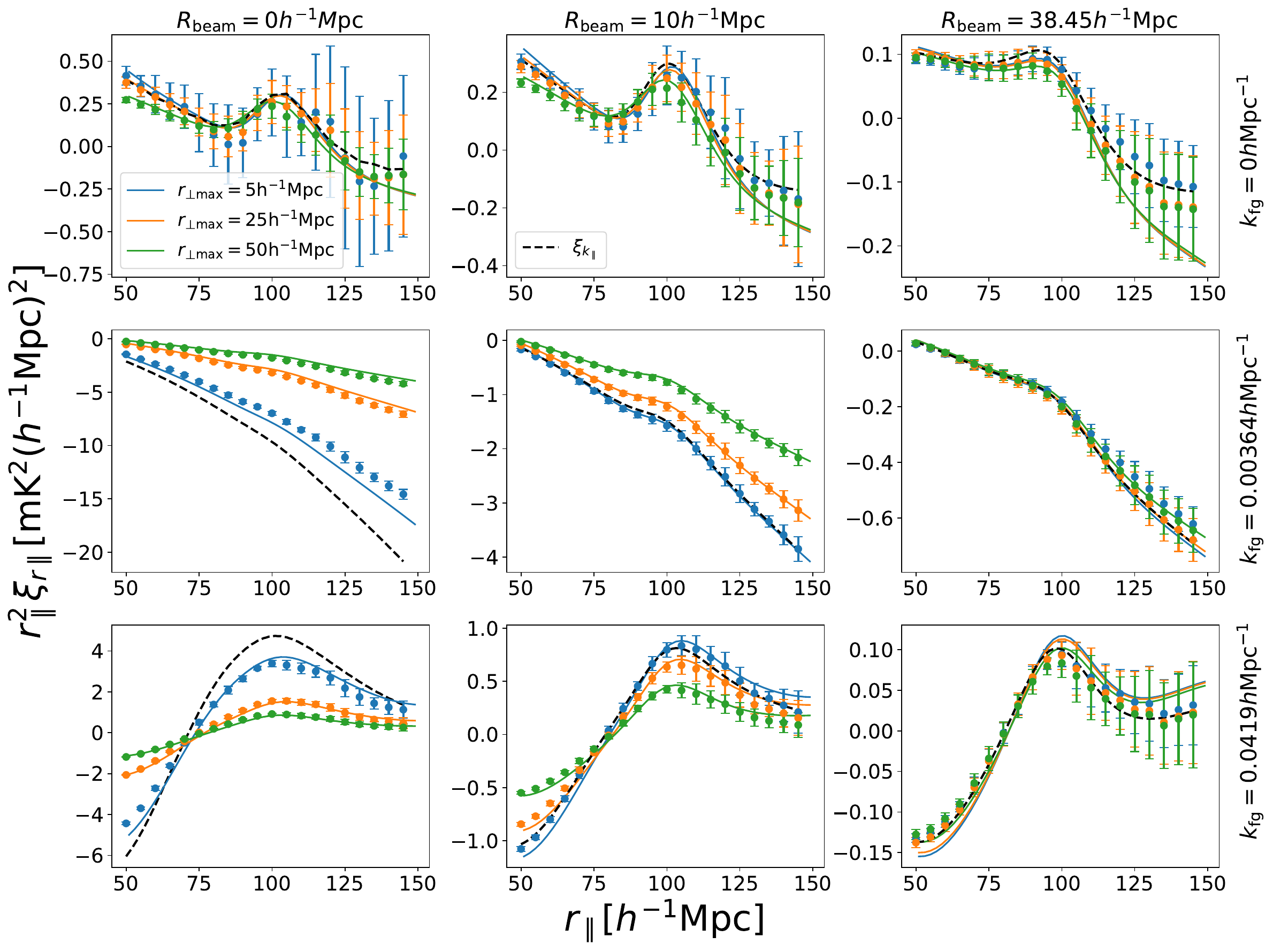}
\caption{ Radial correlation function $\xi_{r \parallel} $ (color, solid) and $\xi_{k \parallel} $ (black, dashed) for different values of $R_{\rm beam}$ (0, 10 and 38.45 $\MpcOh$ from left to right) and $k_{\rm fg}$ (0, 0.00364, 0.0419 $\hOMpc$ from top to bottom). The simulation measurements (markers, the error bars correspond to the standard deviation among the twelve realizations available) and the model (Eq.~\eqref{eq:xi_rparallel}, solid lines) are compared. We have shown $\xi_{r \parallel} $ obtained with $r_{\perp \rm max}  = 5 $ (blue), 25 (orange), and 50 (green) $\MpcOh$, respectively.   }
\label{fig:radial_xiparallel}
\end{figure*}

\subsection{21 cm correlation function model }

The correlation function follows from the power spectrum by an inverse Fourier transform:
\begin{equation}
  \label{eq:xi21cm_full}
  \xi_{\rm 21 cm}(r_{\parallel},r_{\bot})= \int \frac{d^3k}{(2\pi)^3} e^{i \bm{k} \cdot \bm{r} }   P_{\rm 21 cm }(k_\parallel, k_\perp ). %e^{-k_{\bot}^{2}R_{\rm beam}^{2}} \big[1-e^{- k_{//}^2 / k_{\rm fg}^2} \big]^2 
\end{equation}
By utilizing the azimuthal symmetry of the power spectrum with respect to the line of sight, Eq.~\eqref{eq:xi21cm_full} becomes  	
\begin{align}
  \xi_{\rm 21 cm}  &(r_{\parallel},r_{\bot}) =b_{\rm 21 cm}^2 \int \frac{dk_{\parallel}}{ 2 \pi }   \big(1-e^{- k_{\parallel}^2 / k_{\rm fg}^2} \big)^2   e^{ik_{\parallel}r_{\parallel}}  \nn   \\
  & \times \int \frac{dk_{\bot}}{ 2\pi } k_{\bot} e^{-k_{\bot}^{2}R_{\rm beam}^{2}} J_{0}(k_{\bot}r_{\bot})P_{\rm m}(k_{\parallel},k_{\bot}), 
  \label{eq:xi21cm}
\end{align}
where $J_0$ is the zeroth-order Bessel function.

	%We use FFT to calculate the first part of the integral, and then use numerical calculations to calculate the second part of the integral.It should be noted that the input frequency order for FFT is 0,1,2...n-1,-n,-(n-1),...-1.The calculation result is shown in Figure \ref{fig:xi2D} .

Fig.~\ref{fig:xi2D_model} showcases the 2D 21 cm IM correlation function $\xi_{\rm 21 cm}$ (boosted by  a factor of $r^2$ to highlight the BAO feature) computed with different $R_{\rm beam}$ (0, 10, and 38.45 $\MpcOh$) and $k_{\rm fg}$ (0, 0.00364, and 0.0419 $\hOMpc$) values.  As $R_{\rm beam}$ increases, the smoothing effect increases and the BAO feature becomes more blurred. Curiously, we find that at the region close to the $r_\perp$ axis, a peak appears, and it shifts to larger $r_\perp$ as $R_{\rm beam}$ increases. To shed light on this feature, we note the following closed-form relation
\beq
\int_0^\infty d k_\perp k_\perp e^{- k_\perp^2 R_{\rm beam}^2 }  J_0(k_\perp r_\perp )
= \frac{ 1} {2 R_{\rm beam}^2 } e^{- r_\perp^2 / ( 4 R_{\rm beam}^2 ) }. 
\eeq
When $R_{\rm beam}$ increases, this function becomes more and more broadly distributed, and hence with the boost of $r^2$, it can manifest as a peak feature.  Although with $ P_{\rm m}$, the  $k_\perp$-integral in Eq.~\eqref{eq:xi21cm} is no longer analytically integrable, this feature is expected to be inherited. Secondly, to understand that the peak is restricted to the $r_\perp$ axis, i.e.,~$r_\parallel = 0 $, we focus on the radial integral 
\beq
\int d k_\parallel e^{i k_\parallel r_\parallel } P\Big( \sqrt{k_\parallel^2 + k_\perp^2} \Big).  
\eeq
We have neglected the $B_{\rm fg}^2$ term because, as evident in Fig.~\ref{fig:xi2D_model}, this does not qualitatively change the feature. Applying the argument that the integral is highly oscillatory and hence large cancellation results except for $r_\parallel =0 $ [see similar argument in \citet{Dodelson_2003}], we conclude that the feature is significant only on the $r_\perp$-axis. Alternatively, from the uncertainty relation, the spread in configuration space is related the width of $P$ in Fourier space, $\Delta k_\parallel$, as $ \Delta r_\parallel \sim 1 / \Delta k_\parallel $.

As $k_{\rm fg} $ increases, color bands show up, predominantly along the $r_\parallel $-axis.  The factor $B_{\rm fg} $ is a high-pass filter, which  keeps the small-scale information and suppresses the large-scale one. Moreover, this cut-off in Fourier space leads to oscillations in the correlation function. The trough band in the case of $k_{\rm fg}=0.00364 \hOMpc$ and the trough and peak for $k_{\rm fg}=0.0419 \hOMpc$ are the consequences of this oscillatory behavior.   The reason that these oscillatory patterns are only apparent close to the $r_\parallel$ region is because $B_{\rm fg} $ only suppresses the modes with $k_\parallel \lesssim k_{\rm fg}$, and this is significant only if $k_\perp$ is also at a similar order, i.e.,~$k_\perp \lesssim k_{\rm fg}$.   Together with the fact that $ J_0(k_\perp r_\perp) $ is significant for modes satisfying  $k_\perp \sim 1/r_\perp $, we conclude that the bands appear around $r_\perp \sim 1/ k_{\rm fg} $.  We indeed see that as $k_{\rm fg}$ increases, the characteristic $r_\perp$ scale of the bands decreases.

\subsection{ Three types of correlation functions }

From Fig.~\ref{fig:xi2D_model}, it is clear that the beam and foreground cleaning introduce strong anisotropies. The isotropic correlation function alone is no longer effective for BAO measurement. To optimize the BAO extraction from the observed 21 cm IM clustering signals, we consider three types of correlation functions tailored to the 21 cm IM anisotropies, as in \citet{Avila_etal2022}.

\begin{figure*}[!htb]
  \includegraphics[width=0.98\linewidth]{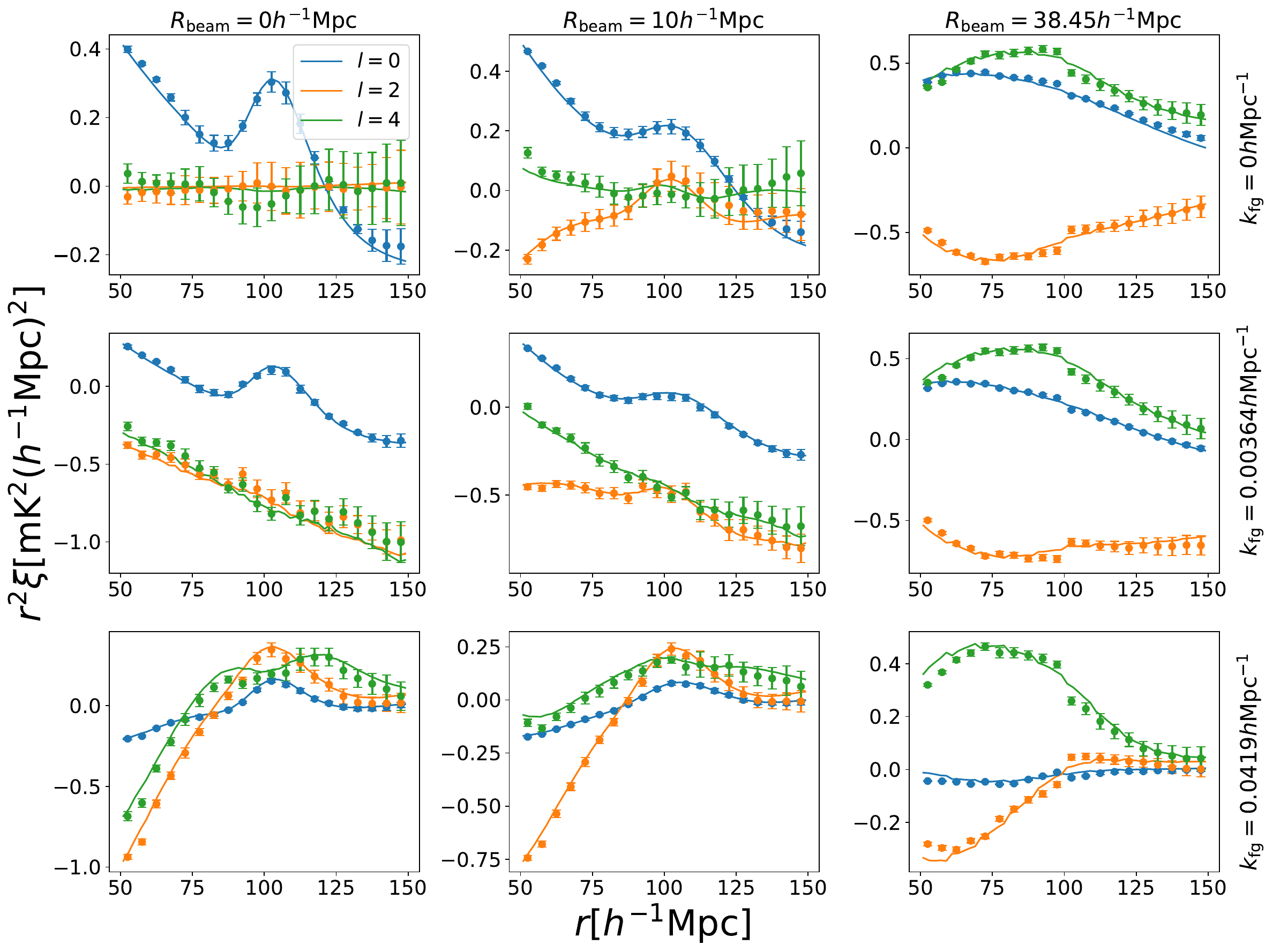}
\caption{ Multipole correlation function $\xi_{l} $ for different values of $R_{\rm beam}$ (0, 10 and 38.45 $\MpcOh$ from left to right) and $k_{\rm fg}$ (0, 0.00364, 0.0419 $\hOMpc$ from top to bottom). The results from simulation measurements (markers; the error bars correspond to the standard deviation among the 12 realizations available) and model (Eq.~\eqref{eq:multipole_correlation}, solid lines) are compared.  Monopole ($l=0$, blue), quadrupole ($l=2$, orange), and hexadecapole ($l=4$, green) results are displayed.  }
\label{fig:multipole_correlation}
\end{figure*}

\subsubsection{Radial correlation function }

To deal with the strong smoothing effect in the transverse direction induced by the beam, \citet{Villaescusa-Navarro_etal2017} proposed the radial power spectrum for 21 cm IM BAO measurement:
\beq
P_\parallel (k_\parallel )= \int \frac{d^2 k_\perp }{(2\pi)^2 } P_{\rm 21cm}(k_\parallel, k_\perp). 
\eeq
By construction, the transverse modes are integrated out, and only the radial ones are left.  
Upon inverse Fourier transform, we get the corresponding radial correlation function 
\begin{align}
\xi_{ k \parallel} (r_\parallel) % &= \int \frac{ d k_\parallel}{2 \pi }  e^{ i k_\parallel r_\parallel } P_\parallel (k_\parallel ) \nn \\
\label{eq:xi_kparallel}
& =  \int \frac{d k_\parallel }{ 2\pi } e^{ i k_\parallel r_\parallel } \int \frac{ d k_\perp}{2 \pi} k_\perp  P_{\rm 21cm}(k_\parallel, k_\perp). 
\end{align}
%In the definition in \citet{Villaescusa-Navarro_etal2017}, there is no explicit cut-off in the integral. However, it is necessary in the case of shot noise as we will see in Sec.~\ref{sec:cov_radial_cf}.   

Alternatively, we can construct a radial correlation function by averaging over the correlation function up to some maximum transverse scale, $r_{\perp \rm max}$  \citep{Avila_etal2022}: 
\beq
\label{eq:radialxi_Avila}
\xi_{ r \parallel}(r_\parallel; r_{\perp \rm max} ) = \frac{1}{ r_{\perp \rm max}^2 } \int_0^{ r_{\perp \rm max} } d r_{\perp } 2 r_\perp \xi_{\rm 21cm } ( r_\parallel, r_\perp ). 
\eeq
Using the relation  $\int \, dx J_0(x) = xJ_1(x)$,  Eq.~\eqref{eq:radialxi_Avila} can be expressed as 
\begin{align}
\xi_{r \parallel}(r_\parallel; r_{\perp \rm max} ) & = \frac{2}{ r_{\perp \rm max } } \int \frac{d k_\parallel}{ 2\pi}  e^{ i k_\parallel r_\parallel} \int \frac{dk_\perp }{2 \pi } \nn \\
\label{eq:xi_rparallel}
& \times J_1(k_\perp r_{\rm \perp max} ) P_{\rm 21 cm } (k_\parallel, k_\perp ),
\end{align}
where $J_1$ is the first-order Bessel function.

One reason against using a very large $r_{\perp \rm max}$ is that if we take $r_{\perp \rm max}$ to infinity, then we get
\begin{align}
\xi_{r \parallel}(r_\parallel) &= \frac{1}{A} \int d^2 r_\perp \xi_{\rm 21cm}(r_\parallel, r_\perp) \nn \\
& = \frac{1}{A} \int \frac{d k_\parallel}{2 \pi} e^{i k_\parallel r_\parallel } P_{\rm 21cm}( k_\parallel, k_\perp = 0 ),
\end{align}
where $A$ is the (irrelevant) transverse area normalization. The key point is that the resultant correlation only depends on the power spectrum  at $k_\perp = 0 $, and this does not seem to be an effective use of information.

On the other hand, from Eqs.~\eqref{eq:xi_kparallel} and \eqref{eq:xi_rparallel}, we find that they are similar in form, with  $\xi_{r \parallel }$ having an additional factor of the jinc function (or sombrero function)
\beq
\mathrm{jinc}( k_\perp r_{\rm \perp max}  ) \equiv \frac{2 J_1(k_\perp r_{\rm \perp max} )}{ k_\perp r_{\rm \perp max}  },
\eeq
which tends to unity as the  argument $k_\perp r_{\rm \perp max} $ approaches 0. This implies that for small $r_{\rm \perp max} $, these two statistics are equivalent.  Thus we can regard $r_{\perp \rm max} $ as a knob controlling the range of $k_\perp $ over which  the power spectrum is averaged.  Taking an intermediate value of $ r_{ \perp \rm  max }  $ can be interpreted as averaging up to some intermediate  $ k_{\rm \perp max} $ scale in Fourier space.

Fig.~\ref{fig:radial_xiparallel} shows the radial correlation function $\xi_{r \parallel}$ for different values of $r_{\perp \rm max}$: 5, 25, and 50 $\MpcOh$. We also show the  $\xi_{k \parallel} $ prediction, which is closest to the   $\xi_{r \parallel} $ result with $ r_{\perp \rm max} = 5 \MpcOh $. The BAO feature appears to be the  strongest for $r_{\perp \rm max} = 5 \MpcOh$. It is important to remember that this takes only the signals into account, but in reality the signal-to-noise ratio (S/N) counts. We shall compare the BAO constraint obtained from different  $r_{\perp \rm max} $ values in Sec.~\ref{sec:optimal_parameter_choices}.  We note that the $k_{\rm fg} = 0.00364 \hOMpc$ cases seem to be distinct from the other two $k_{\rm fg}$ cases. Inspection of Fig.~\ref{fig:xi2D_model} reveals that this is caused by the strong dark band, which suppresses the correlation function around the BAO scale.

%Moreover, as $r_{\perp \rm max} \MpcOh$ increases, the BAO peak shifts to smaller scales relative to the fiducial BAO scale.  This is because although there is suppression of the transverse modes by $B_{\rm beam}$, some residual signals remain and they cause a shift in BAO scale of the radial correlation function. By restricting to a smaller  $r_{\perp \rm max} $, this problem  can be alleviated. This is another reason for favoring a small $r_{\perp \rm max} $.

\begin{figure*}[!htb]
  \includegraphics[width=0.98\linewidth]{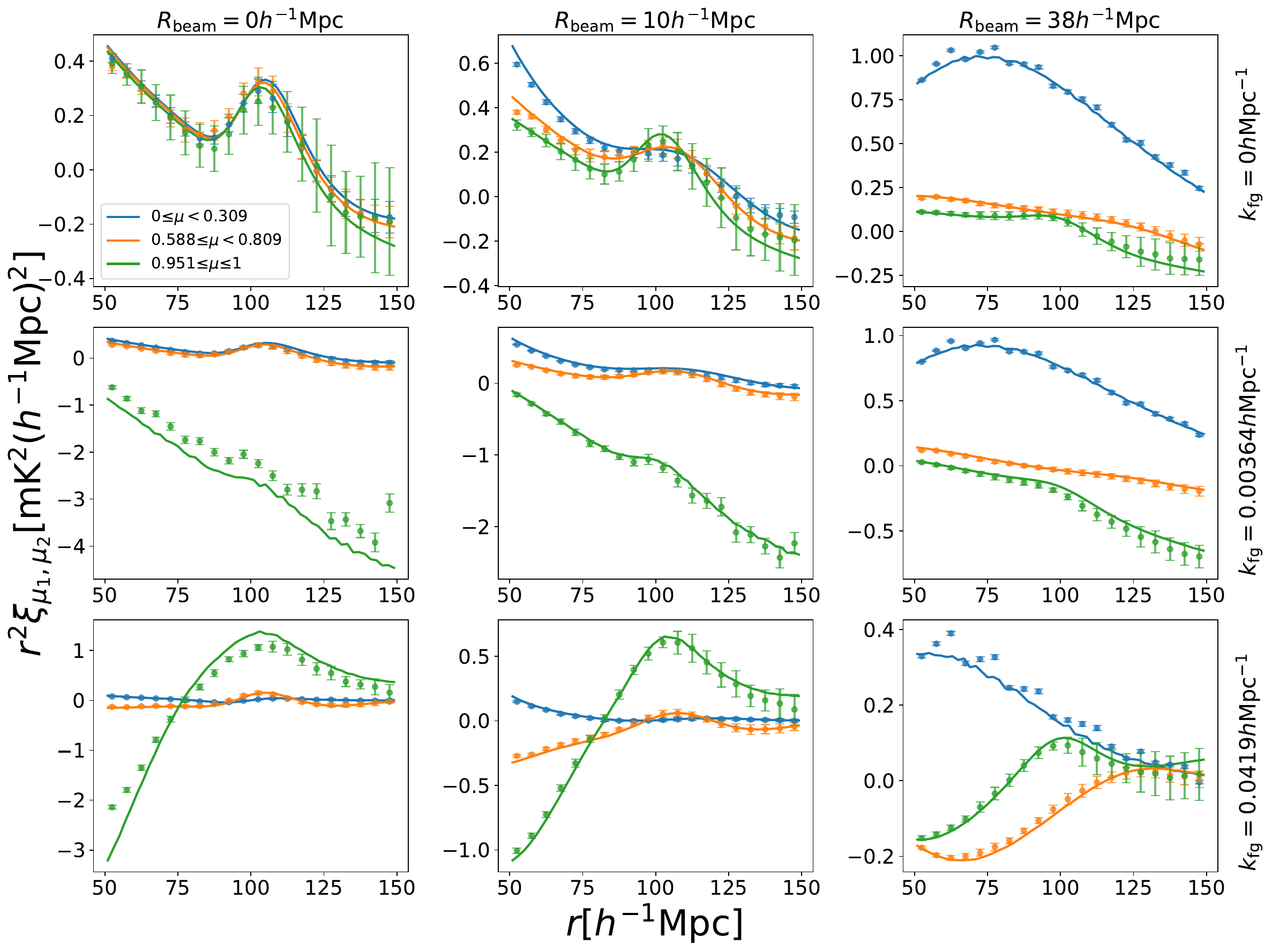}
\caption{ $\mu$-wedge correlation function $\xi_{\mu_1, \mu_2} $ for different  values of $R_{\rm beam}$ (0, 10 and 38.45 $\MpcOh$ from left to right) and $k_{\rm fg}$ (0, 0.00364, 0.0419 $\hOMpc$ from top to bottom). The simulation measurements (markers, the error bars correspond to the standard deviation among the 12 realizations available) and model (Eq.~\eqref{eq:muwedge_correlation}, solid lines) are compared. The angular range [0, 90]$^\circ$ is divided into five bins with equal angular width, and for clarity, we only show the results for the first (blue), third (orange), and last (green) bins.     }
\label{fig:wedge_correlation}
\end{figure*}

\subsubsection{ Multipole  correlation function }

While the radial correlation function keeps only the radial degrees of freedom, the multipole correlation functions, in principle, can retain the full anisotropic clustering information. The multipole correlation function is defined as 
\beq
\label{eq:multipole_correlation}
\xi_{l}(r)=	\frac{2l+1}{2} \int_{-1}^1 d \mu \, \xi_{\rm 21cm} (r,\mu)  \mathcal{L}_l (\mu) , 
\eeq
where $\mu \equiv r_\parallel /r $ and $ \mathcal{L}_l$ represents the $l$th-order Legendre polynomial. 
 Conversely, the correlation function can be built from the multipoles as 
\beq
\xi_{\rm 21cm } (r,\mu) = \sum_{l=0}^\infty \xi_l (r)   \mathcal{L}_l (\mu). 
\eeq
The multipole correlations are commonly employed in the anisotropic galaxy clustering, in which the anisotropies are caused by the RSD and the Alcock-Paczynski effect, and up to the quadrupole is often sufficient to capture the anisotropic information.  Because of the strong anisotropies due to the telescope beam and foreground removal effect,  \citet{KennedyBull_2021} advocated the usage of the multipole correlation function for 21 cm IM BAO measurement.

In Fig.~\ref{fig:multipole_correlation}, we show the monopole ($l=0$), quadrupole ($l=2$), and hexadecapole ($l=4$) correlation functions for various values of $R_{\rm beam}$ and $k_{\rm fg}$. When $ R_{\rm beam} =0 \, \MpcOh$ and $k_{\rm fg}=0 \, \hOMpc$, i.e.,~the isotropic case, all the BAO information is concentrated in the monopole. When the values of $R_{\rm beam}$ and $k_{\rm fg}$ are cranked up, some BAO information passes to the higher-order multipoles. The BAO signals become progressively smaller as $l$ increases.   When $R_{\rm beam }$ reaches $38.45 \MpcOh $, the BAO signature is barely visible in the multipoles.

\subsubsection{$\mu$-wedge  correlation function }

Like the multipole correlation function, the $\mu$-wedge correlation function is also capable of retaining the full anisotropic information and is often employed in anisotropic galaxy clustering analysis \citep{Sanchez_etal2013}.  In this case,  it allows the avoidance of the strong distortion caused by the Finger-of-God along the line-of-sight direction in redshift space.  To bypass the large distortion along and transverse to the line of sight direction in  $\xi_{\rm 21cm }$, \citet{Avila_etal2022} proposed the $\mu$-wedge correlation function 
\begin{equation}
\label{eq:muwedge_correlation}
\xi_{\mu_{1},\, \mu_{2}}(r) = \frac{1}{\mu_{2} - \mu_{1} }  \int_{ \mu_{1} }^{ \mu_{2} } d \mu \, \xi_{\rm 21cm} (r,\mu),
\end{equation}
to characterize the anisotropic correlation function.  The limits $\mu_{1}$ and $\mu_{2}$ were chosen to avoid the features due to the beam and foreground removal effect. In general, these non-cosmological artifacts are hard to model; a simple strategy is to avoid them altogether. To enhance the S/N, we will consider the $\mu$-wedge correlation function refined into  a number of bins.

We plot the $\mu$-wedge correlation function in Fig.~\ref{fig:wedge_correlation}. We have divided the quadrant [0,90]$^\circ$ into five bins, each of equal angular bin width. For clarity, we only show the first, third, and last bin results.  We note that for the first and last bins, the models are often not in very good agreement with the measurement. This is due to the artifacts caused by the beam and foreground removal effects, and they are localized in the region close to $\mu \sim 0$  and  $\sim 1$, respectively. We will consider getting rid of these problematic regions in the fitting.

\section{ Correlation function covariances } 
\label{sec:covariances}

To perform the BAO fit, besides the correlation function model, we also require its covariance.   For a correlation function estimator $\hat{\xi}(\bm{r})$, its covariance is defined as 
\begin{equation}
  \label{eq:general_3Dcov}
   	   C(\bm{r},\bm{r'})=\left \langle \hat{\xi}(\bm{r}) \hat{\xi}(\bm{r}')\right \rangle - \left \langle \hat{\xi}(\bm{r})\right \rangle \left \langle \hat{\xi}(\bm{r}')\right \rangle . 
   \end{equation}
In this work, we adopt the Gaussian theory covariance. \change{ Gaussian covariance is commonly used in BAO fitting, especially in forecast studies. It has been demonstrated to be in good agreement with mock results \citep{Grieb_etal2016,Chan_etal2018,DES_Y6BAO2024,Rashkovetskyi_etal2025}, e.g., the generalized Gaussian covariance yields less than an 8\% difference in the error bars of the distance measurements relative to mock results in \citet{Rashkovetskyi_etal2025}. }

The Gaussian covariance for $\xi_{\rm 21 cm} $ is given by
\begin{align}
  \label{eq:covmat_21cm_full}
C_{\rm 21 cm} ( \bm{r},\bm{r'}) & = \frac{1}{V} \int \frac{d^3 k }{(2 \pi)^3} \Big(  P_{\rm 21cm }( \bm{k} )  + P_{\rm N}  \Big)^2 \nn \\
& \times \big[  e^{ i \bm{k} \cdot ( \bm{r} - \bm{r}' ) } +    e^{ i \bm{k} \cdot ( \bm{r} + \bm{r}' ) }    \big],   
\end{align}
where $V$ either is the simulation volume or the survey volume.  In the case of galaxy clustering, $ P_{\rm N}  $ usually represents the Poisson shot noise, and it can be written in terms of the number density of the galaxy sample, $\bar{n} $ as $P_{\rm N} =  1 / \bar{n} $.  For 21 cm IM clustering, the instrumental noise is the main source of noise.  We can absorb this white noise contribution into $P_{\rm N}$  using an effective $\bar{n} $ (or $\bar{n}/ \bar{T}_{\rm b}^2 $ because in the convention of radio clustering, the power spectrum carries an additional factor of  $\bar{T}_{\rm b}^2 $) \citep{Villaescusa-Navarro_etal2017,KennedyBull_2021}. The Poisson noise due to the system temperature $T_{\rm sys}$  is given by
\beq
\label{eq:PN_radiometereq}
P_{\rm N} =  T_{\rm sys}^2  \frac{ A_{\rm sky} \lambda_{\rm 21cm} [ (1+z) r ]^2   }{ 2 N_{\rm dish} t_{\rm tot}   H(z)   }, 
\eeq
where  $A_{\rm sky} $ is the sky area in steradian, $c$ is the light speed, $t_{\rm tot}$ is the total integration time,  $ \lambda_{\rm 21cm} $ is the wavelength of the rest-frame 21 cm signal, and $ N_{\rm dish} $ is the number of dishes. This additional factor of 2 in the denominator is due to the two polarizations assumed. This white noise contribution can be derived from radiometer equation \citep{Thompson_etal2001}, see, for example,~\citet{Villaescusa-Navarro_etal2017}.  For the tests in the first part, we use a fixed value of $ P_{\rm N} = 30 \, \mathrm{ (mK)^2} (\MpcOh)^3 $, which is on the order of magnitude expected in usual 21 cm experiments, and in the second forecast part, the value is computed based on the survey specifications.

In the following, we present the Gaussian covariance for the three types of correlation functions introduced. These covariances can be viewed as different filtering or projection of the full covariance (Eq.~\eqref{eq:covmat_21cm_full}).

\subsection{ Radial correlation function covariance }
\label{sec:cov_radial_cf}

We first derive the  $\xi_{k \parallel} $ covariance and then discuss the  $\xi_{r \parallel} $ covariance at the end.  Plugging in the estimator for  $\xi_{k \parallel } $
\begin{align}
\hat{\xi}_{k \parallel} (r_\parallel) = \int \frac{d k_\parallel  }{2 \pi} e^{i k_\parallel r_\parallel}  \int \frac{ d^2 k_\perp}{(2\pi)^2} \frac{\delta_{\rm 21cm}(\bm{k} ) \delta_{\rm 21cm}(-\bm{k} )  }{V}, 
\end{align}
with $\delta_{\rm 21 cm}$ being the 21 cm IM fluctuation in Fourier space,
it is easy to  write down its Gaussian covariance: 
\begin{align}
   	  C_{k \parallel}(r_{\parallel},r'_{ \parallel}) &  = 
       \frac{2}{V} \int \frac{dk_{\parallel}}{2\pi} e^{ik_{\parallel} (r_{\parallel} - r_{\parallel}' )  } \int \frac{d^2k_{\bot}}{(2\pi)^2} \nn \\  & \times \Big( P_{\rm 21cm}(k_\parallel, k_\perp )+ P_{\rm N}  \Big)^2 .
\end{align}

In the computation of covariance, it is necessary to account for the fact that the measured value is averaged within a finite bin width.  We consider a bin  centering at  $ r_\parallel $ with width $\Delta$, i.e.,~$[ r_\parallel - \Delta  /2,  r_\parallel + \Delta /2 ] $.  We first notice that 
\beq
\frac{1}{ \Delta }  \int_{r_\parallel - \Delta /2}^{r_\parallel + \Delta /2}  d  \rho \, e^{ ik_\parallel \rho}  = e^{i k_\parallel r_\parallel } \mathrm{sinc} \Big( \frac{ k_\parallel \Delta }{2} \Big), 
\eeq
where $ \mathrm{sinc} x \equiv \sin x / x$.  Thus, by averaging over the $r_\parallel$ and $r'_\parallel$ bins, we have  
\begin{align}
   	  C_{k \parallel}(r_{\parallel},r'_{\parallel}) &  = 
       \frac{2}{V} \int \frac{dk_{\parallel}}{2\pi} e^{ik_{\parallel} (r_{\parallel} - r_{\parallel}' )  }   \mathrm{sinc}^2 \Big(\frac{ k_\parallel \Delta }{2} \Big)   \nn \\   
      & \times \int \frac{d^2k_{\bot}}{(2\pi)^2} 
        \Big( P_{\rm 21cm}(k_\parallel, k_\perp )+ P_{\rm N}   \Big)^2 . 
\end{align}
The results can be broken down into 
\begin{align}
C_{k \parallel}(r_\parallel , r'_\parallel) = \frac{2}{V} \Big(  S_1 +  2 P_{\rm N} S_2  + P_{\rm N}^2  S_3 \Big) , 
\end{align}
where $S_1$, $S_2$, and $S_3$ denote 
\begin{align}
 S_1   = b_{\rm 21cm}^4   \int \frac{ d k_\parallel }{2 \pi}  & e^{i k_\parallel ( r_\parallel - r'_\parallel ) }   \mathrm{sinc}^2 \frac{ k_\parallel \Delta }{2}  B_{\rm fg}^4 (k_\parallel )  
  \int \frac{ d k_\perp \, k_\perp  }{2 \pi}   \nn \\ 
 &  \times B^4_{\rm beam} (k_\perp ) P_{\rm m }^2 ( k_\parallel, k_\perp),  \\
 S_2   = b_{\rm 21cm}^2   \int \frac{ d k_\parallel }{2 \pi}  & e^{i k_\parallel ( r_\parallel - r'_\parallel ) }   \mathrm{sinc}^2 \frac{ k_\parallel \Delta }{2}  B_{\rm fg}^2 (k_\parallel )  \int \frac{ d k_\perp \, k_\perp  }{2 \pi}   \nn \\
  &\times B^2_{\rm beam} (k_\perp ) P_{\rm m } ( k_\parallel, k_\perp),  \\
  \label{eq:Ckparallel_S3}
 S_3   = \frac{k_{\rm \perp \rm max}^2 }{4 \pi}  & \int \frac{ d k_\parallel }{2 \pi} e^{i k_\parallel ( r_\parallel - r'_\parallel ) }   \mathrm{sinc}^2 \frac{ k_\parallel \Delta }{2}  .
\end{align}
In $S_3$, $k_{\rm \perp \rm max}$ is the cut-off introduced in Fourier space in order to make the integral finite.  
%\KCC{The radial power spectrum is ill-defined if there is no cut-off in perpendicular integral. For shot noise, the integral diverges. In the covariance expression in \citet{Villaescusa-Navarro_etal2017}, it suddenly introduced a cut-off in the covariance in Eq.C5 there. Let's say there is a maximum cut-off scale in the definition.    }

\begin{figure}[!t]
  \centering
    \includegraphics[width=\linewidth]{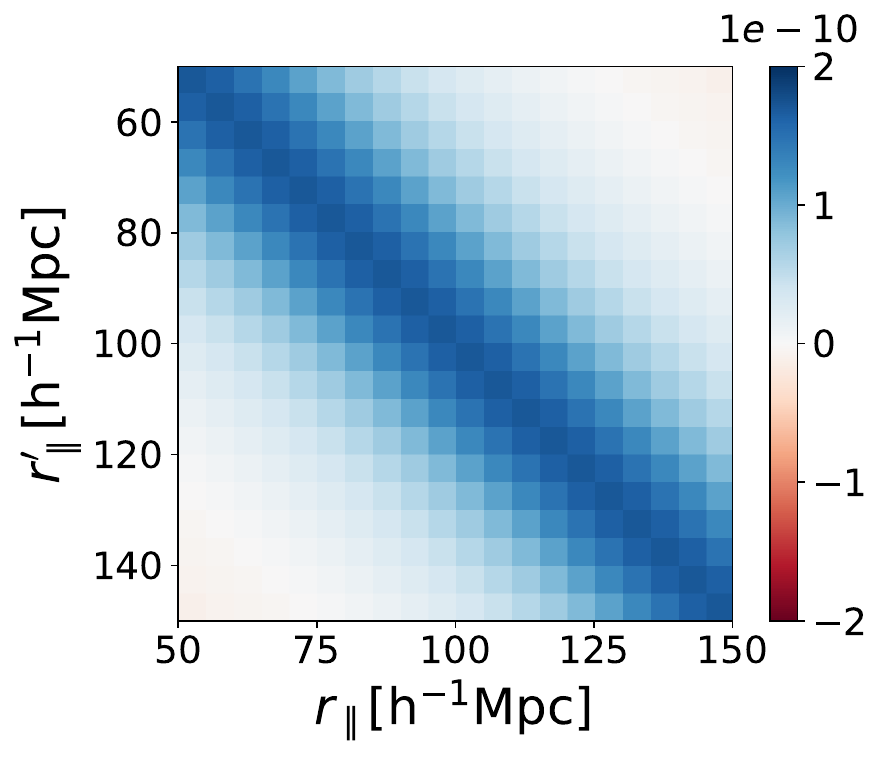}
  \caption{ Radial correlation function covariance with $ R_{\rm beam} = 10 \, \MpcOh $, $ k_{\rm fg} = 0.0036 \, \hOMpc $, and  $ r_{\perp \rm max} = 50 \MpcOh $.    }
  \label{fig:radial_covmat}
\end{figure}

We now turn to the covariance of $\xi_{r \parallel} $. Following the earlier comment that $\xi_{r \parallel }  $ can be derived from  $\xi_{k \parallel }  $ by introducing an additional factor of
 $   \mathrm{jinc}( k_\perp r_{\rm \perp max}  ) $, we can immediately write down the corresponding covariance for $\xi_{r \parallel} $:
\begin{align}
   	 C_{r \parallel}(r_{\parallel},r'_{\parallel})  &  = 
       \frac{2}{V} \int \frac{dk_{\parallel}}{2\pi} e^{ik_{\parallel} (r_{\parallel} - r_{\parallel}' )  }   \mathrm{sinc}^2 \Big(\frac{ k_\parallel \Delta }{2} \Big)     
  \int \frac{d^2k_{\bot}}{(2\pi)^2}   \nn \\  
  & \times   \mathrm{jinc}^2( k_\perp r_{\rm \perp max}  ) \Big( P_{\rm 21cm}(k_\parallel, k_\perp )+ P_{\rm N}  \Big)^2 . 
\end{align}
We show a plot of the radial correlation function covariance in Fig.~\ref{fig:radial_covmat}. In this plot, we have taken  $ R_{\rm beam} = 10 \, \MpcOh $, $ k_{\rm fg} = 0.0036 \, \hOMpc $, and  $ r_{\perp \rm max} = 50 \MpcOh $.

  \subsection{ Multipole correlation function covariance }

The covariance for the multipole correlation function can be written in terms of the full covariance $C( \bm{r},\bm{r'})$ (Eq.~\eqref{eq:covmat_21cm_full}) as 
\begin{align}
\label{eq:Cmultipole_original}
	   C_{ll'}(r, r') & =\frac{(2l+1)(2l'+1)}{4} \int^{1}_{-1} d\mu \int^{1}_{-1}  d\mu'   \nn \\
      &  \times \mathcal{L}_l(\mu) \mathcal{L}_{l'}(\mu') C( \bm{r},\bm{r'}).  
   \end{align}

Employing the identity 
\beq
\int \frac{ d \hat{ \bm{r} }} {4 \pi } \mathcal{L}_l ( \hat{ \bm{r}}  \cdot \hat{\bm{n}} ) e^{ i \bm{k} \cdot \bm{r}  }  =  i^l j_l(kr) \mathcal{L}_l ( \hat{\bm{n} } \cdot \hat{\bm{k}} ),  
\eeq
we can express Eq.~\eqref{eq:Cmultipole_original} as 
% \begin{align}
%   C_{ll'}(r, r') & =  \frac{1}{V} ( 2 l+1) (2l'+1) i^{l+l'} \big[ 1+ ( -1)^{l'}  \big] \int \frac{d k k^2}{ (2 \pi)^2} j_l(kr) j_{l'} ( kr')  \nn \\
% & \times \int  \frac{d \hat{\bm{k}} }{ 2 \pi}  \mathcal{L}_l (\hat{\bm{k}} \cdot \hat{\bm{n}} )  \mathcal{L}_{l'}  (\hat{\bm{k}} \cdot \hat{\bm{n}}   )    \Big(   P_{\rm 21cm}(k_\parallel, k_\perp ) + \frac{1}{\bar{n} }  \Big)^2  , 
% \end{align}
% where  $\hat{\bm{n}} $ denotes the line of sight direction. Because we consider even multipoles only, then it reduces to 
\begin{align}
  C_{ll'}(r, r') & =  \frac{2}{V} ( 2 l+1) (2l'+1) i^{l+l'} \int \frac{d k k^2}{ (2 \pi)^2} j_l(kr) j_{l'} ( kr')  \nn \\
& \times \int  \frac{d \hat{\bm{k}} }{ 2 \pi}  \mathcal{L}_l (\hat{\bm{k}} \cdot \hat{\bm{n}} )  \mathcal{L}_{l'}  (\hat{\bm{k}} \cdot \hat{\bm{n}}   )    \Big(   P_{\rm 21cm}(k_\parallel, k_\perp ) + P_{\rm N}  \Big)^2 , 
\end{align}
where  $\hat{\bm{n}} $ denotes the line-of-sight direction and we have assumed even multipoles.    Plugging in Eq.~\eqref{eq:Pk_21cmmodel} for $ P_{\rm 21cm} $, the final expression for the multipole covariance reads \citep{Tansella_etal2018, KennedyBull_2021}
\beq
\label{eq:Cllp_covmat}
 C_{ll'}(r, r') = \frac{2}{V} \Big( T_1 + P_{\rm N} T_2 + P_{\rm N}^2 T_3 \Big), 
\eeq
where these three terms are given by 
\begin{align}
T_1 & =  (2l+1)(2l'+1) i^{l+l'}  b_{\rm 21 cm}^4   \int \frac{ dk \, k^2}{  (2 \pi)^2}j_l(kr)j_{l'}(kr') P_{\rm m}^2 (k) \nn \\
& \times \int d \mu_k \mathcal{L}_l(\mu_k ) \mathcal{L}_{l'}( \mu_k) B_{\rm beam}^4(k_\perp ) B^4_{\rm fg}( k_\parallel ) ,    \\
T_2 & = (2l+1)(2l'+1) i^{l+l'}  b_{\rm 21 cm}^2    \int \frac{ dk \, k^2}{  (2 \pi)^2}j_l(kr)j_{l'}(kr') P_{\rm m} (k) \nn \\
& \times \int d \mu_k \mathcal{L}_l(\mu_k ) \mathcal{L}_{l'}( \mu_k) B_{\rm beam}^2(k_\perp ) B^2_{\rm fg}( k_\parallel ),     \\
T_3 & =  (2l+1)(2l'+1) i^{l+l'} \int \frac{ dk \, k^2}{  ( 2 \pi )^2}j_l(kr) j_{l'}(kr')  \nn \\
& \times \int d \mu_k \mathcal{L}_l(\mu_k ) \mathcal{L}_{l'}( \mu_k) . 
\end{align}

  The pure shot noise term, $ T_3$ can be computed analytically. To do so, we first note that 
\beq
\mathcal{L}_l(\mu_k ) \mathcal{L}_{l'}( \mu_k) = \sum_{ L=|l-l'| }^{ l+l' }  \begin{pmatrix}
l & l' & L \\
0 & 0 & 0
\end{pmatrix}^2 (2L+1) \mathcal{L}_{L}( \mu_k) ,
\eeq
where the big brackets denote the Wigner 3-$j$ symbol.  We can then simplify the angular integral in $T_3$ as  
\beqa
   \int d \mu_{k}  \mathcal{L}_l(\mu_k ) \mathcal{L}_{l'}( \mu_k)   
= &&  \sum_{ L=|l-l'| }^{ l+l' } 
\begin{pmatrix}
l & l' & L \\
0 & 0 & 0
\end{pmatrix}^2 (2L+1)  
\int d \mu_{k}    \mathcal{L}_{L}( \mu_k)  \nn \\
= &&  \sum_{ L=|l-l'| }^{ l+l' } 
\begin{pmatrix}
l & l' & L \\
0 & 0 & 0
\end{pmatrix}^2 (2L+1)  
2 \delta_{L0}    \nn \\
= && \frac{ 2 } {2 l+1} \delta_{l l'} .
\eeqa
Moreover, we have
\beq
\int \frac{ dk \, k^2}{  ( 2 \pi )^2 }j_l(kr) j_{l'}(kr') = \frac{\Ddel(r-r')}{8\pi r^2}.  
\eeq
Consequently, $T_3 $ can be reduced to 
\beqa
T_3 = \frac{2l+1}{ 4 \pi  r^2  } \delta_{ll'} \Ddel( r -r'),
\eeqa
for even multipoles.

As in the radial correlation function case, we have to  average over the bins $[ r-\Delta/2, r+\Delta/2 ] $  and  $[ r'-\Delta/2, r'+\Delta/2 ] $. For $T_1$ and $T_2$, this can be done by replacing $j_l$ with its bin-averaged version $\bar{j}_l$:
\begin{align}
\bar{j}_l(r) &= \frac{1}{V(r,\Delta) } \int_{r- \frac{\Delta}{2} }^{r + \frac{\Delta}{2} } d \rho \, 4 \pi \rho^2 j_l(\rho) ,
\end{align}
where $V(r,\Delta) $ is the volume of the shell
\beq
V(r,\Delta) = 4 \pi \Delta  \Big( r^2 +   \frac{\Delta^2}{12} \Big). 
\eeq
Upon bin-averaging, $ T_3  $ is reduced to 
\beq
T_3 = \frac{2l+1}{ 4 \pi r^2 } \delta_{ll'} \frac{\delta_{rr'} }{\Delta}. 
\eeq
This result agrees with \citet{KennedyBull_2021}, although they did not phrase it as bin-averaging and no bin-averaging was applied in $T_1$ or $T_2$.  

\begin{figure}[!t]
  \centering
  \includegraphics[width=\linewidth]{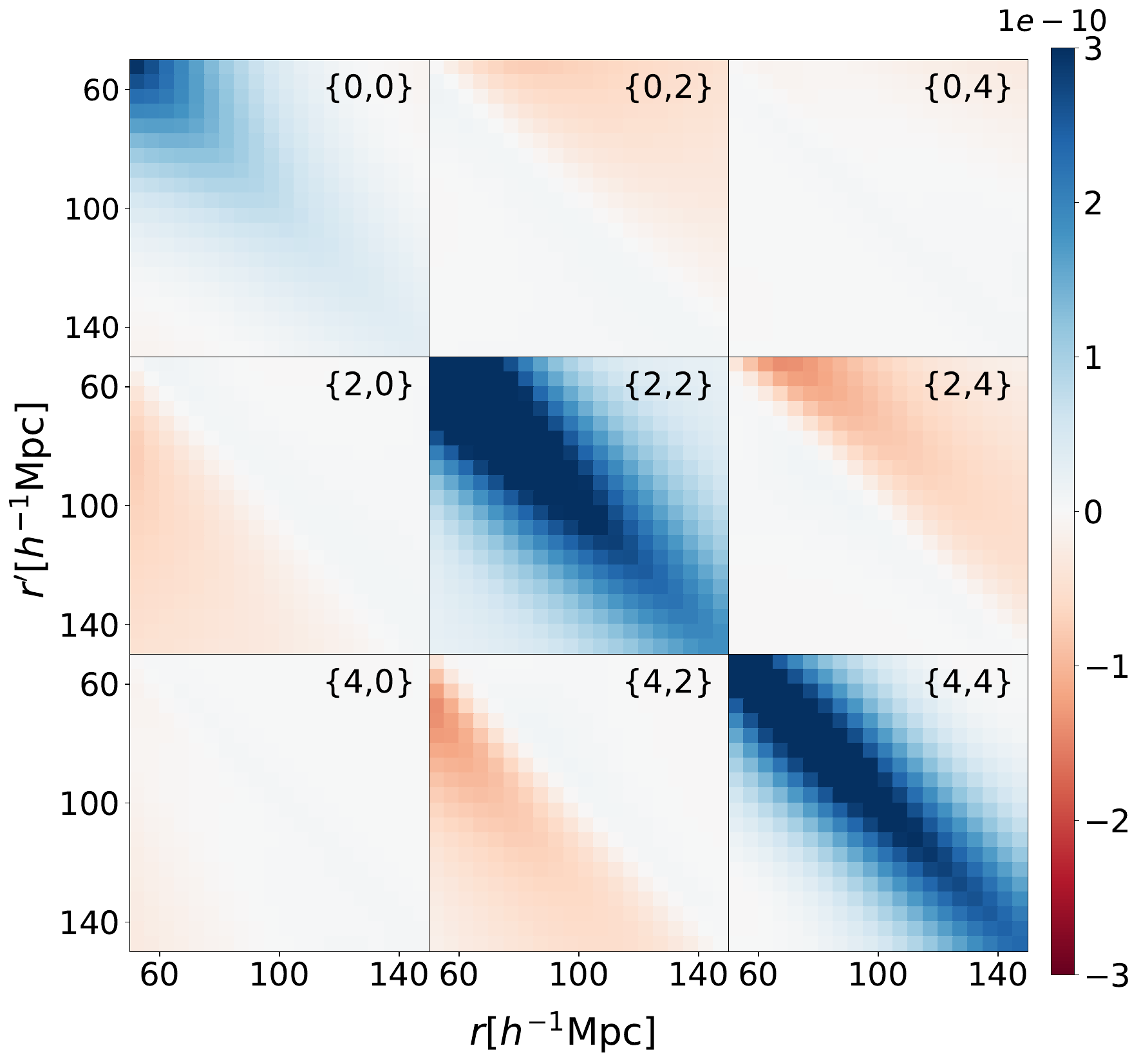}
  \caption{ Multipole correlation function covariance for the monopole ($l=0$), quadrupole ($l=2$), and hexadecapole ($l=4$).  In this plot, we take $ R_{\rm beam} = 10 \, \MpcOh $ and $ k_{\rm fg} = 0.0036 \, \hOMpc $.   }
  \label{fig:multipole_covmat}
\end{figure}

In Fig.~\ref{fig:multipole_covmat}, we plot the multipole correlation function covariance for monopole, quadrupole, and hexadecapole. In this plot, we take $ R_{\rm beam} = 10 \, \MpcOh $ and $ k_{\rm fg} = 0.0036 \, \hOMpc $. The covariance of the quadrupole and hexadecapole is substantially stronger than the monopole one.

   \subsection{ $\mu$-wedge correlation function covariance}

To compute the covariance of the $\mu$-wedge correlation function, we first note that  $\xi_{\mu_0, \mu_1} (r) $ can be expanded in terms of the multipole correlation function as 
\beq
\xi_{\mu_1, \mu_2 }(r) = \sum_l \xi_l (r) \bar{\mathcal{L}}_l (\mu_1, \mu_2) , 
\eeq
where $\bar{\mathcal{L}}_l (\mu_1, \mu_2) $ represents the $\mu$-averaged Legendre polynomial defined as 
\beqa
\bar{\mathcal{L}}_l ( \mu_1, \mu_2 ) & = & \frac{1}{ \mu_2 - \mu_1 } \int_{\mu_1}^{\mu_2}  d \mu  \mathcal{L}_l (\mu) ,
\eeqa
which can be simplified to
\begin{equation}
\bar{\mathcal{L}}_l ( \mu_1, \mu_2 )  = 
\begin{cases}
    \frac{ \mathcal{L}_{l+1}(\mu_2) -  \mathcal{L}_{l-1}(\mu_2)  -  \mathcal{L}_{l+1}(\mu_1) + \mathcal{L}_{l-1}(\mu_1) }{ (2l+1) ( \mu_2 - \mu_1 )}   & \text{ if }  l \ge 1, \\
    1 & \text{ if } l = 0 .
\end{cases}
\end{equation}
 The covariance of the $\mu$-wedge correlation function then reads 
\beq
C \big( \xi_{\mu_1, \mu_2 }(r) , \xi_{\mu_1', \mu_2' }(r') \big) = \sum_{l,l'}  \bar{\mathcal{L}}_l ( \mu_1, \mu_2 )  \bar{\mathcal{L}}_{l'} ( \mu_1', \mu_2' )  C_{ll'}(r,r'),
\eeq
where $C_{ll'}$ is given by the $r$-bin-averaged version of Eq.~\eqref{eq:Cllp_covmat}.  We use up to $l=30$, although for the coarse wedge setting, much fewer multipoles are required.

\begin{figure}[!t]
  \centering
  \includegraphics[width=\linewidth]{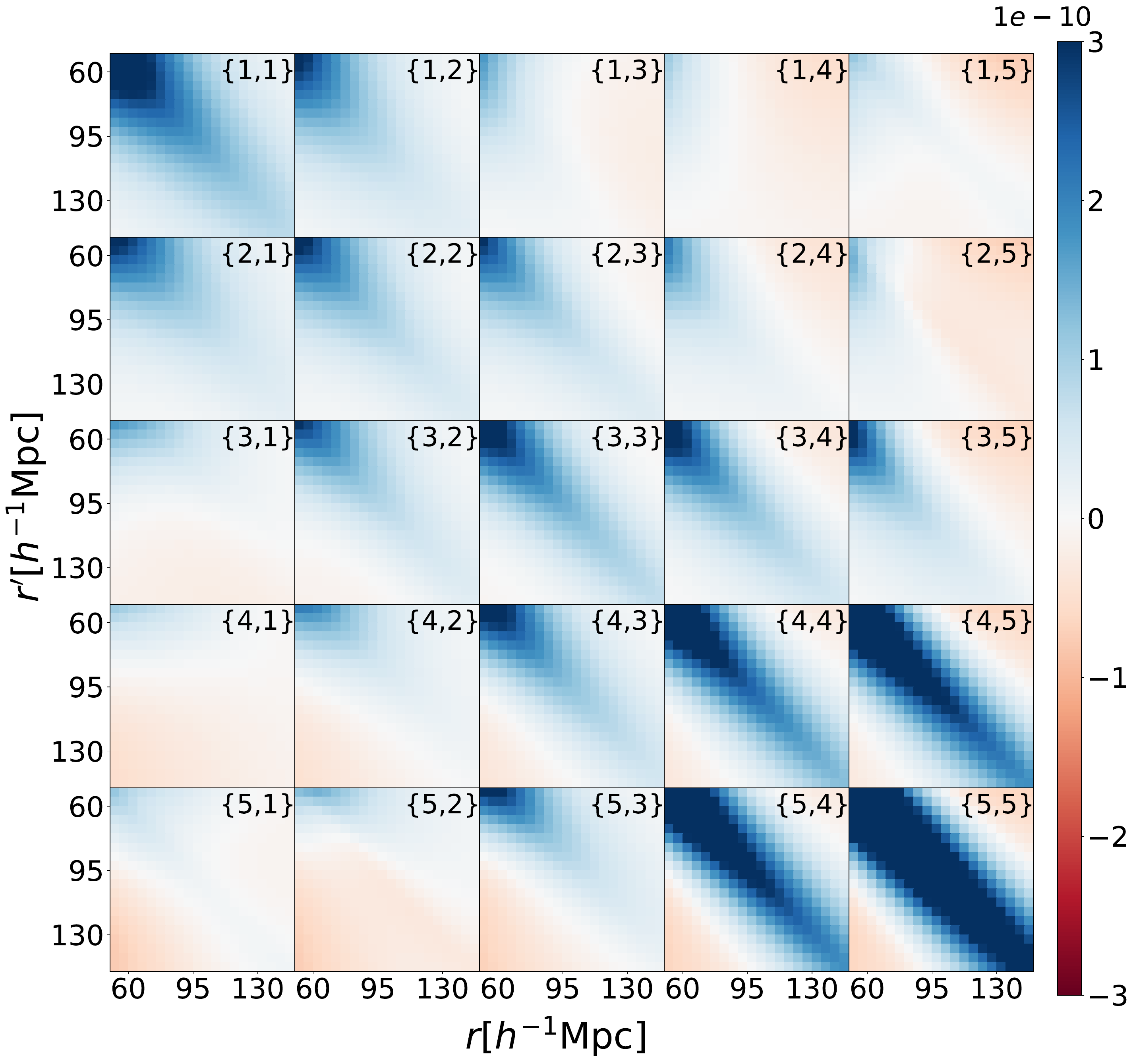}
  \caption{ $\mu$-wedge correlation function covariance for the five-wedge case. 
In this plot, $ R_{\rm beam} = 10 \, \MpcOh $ and $ k_{\rm fg} = 0.00364 \, \hOMpc $ are used.  }
  \label{fig:muwedge_covmat}
\end{figure}

In Fig.~\ref{fig:muwedge_covmat}, we show the $\mu$-wedge covariance with $ R_{\rm beam} = 10 \, \MpcOh $ and $ k_{\rm fg} = 0.00364 \, \hOMpc $. As in Fig.~\ref{fig:wedge_correlation}, the results come from a five-wedge configuration with each wedge of equal angular width. Note that there is strong cross covariance between the neighboring bins. Moreover, the covariance of the first bin and the last two bins is stronger, which is caused by the beam and foreground removal effects, respectively.

\section{ Method }
\label{sec:method}

In this section, we first introduce the \HI simulation used to verify our BAO fitting pipeline and then describe the method  for parameter inference.

\subsection{H$_{\rm I}$ simulation and its correlation function estimation}
\label{sec:HI_similation}

We utilize the H$_{\rm I}$ simulation from \citet{Avila_etal2022}. Here we present a brief overview of the simulation and refer the readers to the original paper for further details.  This simulation set is built upon the UNIT $N$-body simulations \citep{Chuang_etal2019}, which are dark-matter only simulations in a flat $\Lambda$CDM cosmology with cosmological parameters $\Omega_{\rm m}=0.309$, $h=0.6674$, $n_{\rm s}=0.967$, and $\sigma_{8}=0.815$. In each simulation, there are $4096^3$ particles in a cubic simulation box of volume 1  $(\mathrm{Gpc} \,h^{-1})^3$, resulting in the particle mass of $1.25 \times 10^9 M_\odot h^{-1} $. These simulations are run with a modified version of {\tt Gadget2} \citep{Springel2005}.   Although there are four simulations only, they are run with the fixed and paired technique \citep{AnguloPontzen2016} to reduce the covariance.  The four simulations at redshift $z=1.321$ are used. Besides, the three axes of the simulation box are rotated to generate three more sets of simulation realizations. In total, we have 12 realizations.

Halos are identified using the phase-space halo finder {\tt ROCKSTAR} \citep{Behroozi_etal2013_rockstar}, and merger trees are built with {\tt CONSISTENTTREES } \citep{Behroozi_etal2013_mergertree} tracking halos across snapshots.  From the merger history and the mass accretion history of the (sub)-halos, the Semi-Analytic Galaxy Evolution ({\tt SAGE}) model \citep{Croton_etal2016} is run in \citet{Knebe_etal2022} to create the baryonic properties and galaxies in the halos. The model has eight free parameters, which are calibrated with a number of relations, and most of them are observational in nature.  We refer the readers to \citet{Croton_etal2016,Knebe_etal2018} for more details on the model calibration.

To model the H$_{\rm I}$ mass in the galaxies, $M_{\rm H_{\rm I} }$, the cold gas mass, $M_{\rm CG}$ computed by {\tt SAGE} is post-processed with the relation
\beq
M_{\rm H_{\rm I}} = f_{\rm H} \frac{ 1 }{ 1 + R_{\rm mol} } M_{\rm CG}, 
\eeq
where $f_{\rm H}$  is the fraction of hydrogen (set to 0.75 using the Big Bang Nucleosynthesis prediction), and $ R_{\rm mol}= M_{\rm H_2} / M_{\rm H_{\rm I} } $ is the mass ratio between molecular hydrogen and the hydrogen atom. In \citet{Avila_etal2022}, it  is taken to be a constant value, $R_{\rm mol}=0.4$ following \citet{Zoldan_etal2017}. We caution that there are still large uncertainties in the treatment of H$_{\rm I}$, see the discussions in \citet{Avila_etal2022}.  Consequently, the brightness temperature contrast  $\Delta T = T- \bar{T}_{\rm b}$ can be computed with the relation  $\Delta T \propto \Delta M_{\rm H_{\rm I} } = M_{\rm H_{\rm I} } - \bar{M}_{\rm H_{\rm I} }   $, using  $M_{\rm H_{\rm I} } $ obtained from {\tt SAGE}.    

%IM (IM) measures the total flux contribution from galaxies falling with the pixel instead of resolving individual galaxies. IM effectively samples the continuous flux distribution.   In IM, the physical observable is

From the temperature fluctuations $\Delta T$, we measure the isotropic correlation function, expressed in a 2D grid as 
\beq
\xi( r_\perp , r_\parallel ) = \langle  \Delta T ( \bm{x}_\perp,  \bm{x}_\parallel )  \Delta T ( \bm{x}_\perp + \bm{r}_\perp,  \bm{x}_\parallel + \bm{r}_\parallel ) \rangle.     
\eeq
We then perform a Fourier transform on $\xi( r_\perp , r_\parallel )$ by applying an FFT on  $r_\parallel$ and a Hankel transform on  $r_\perp$.  In Fourier space, we add the beam effect (Eq.~\eqref{eq:Rbeam_model}) and foreground removal effect (Eq.~\eqref{eq:Bfg_model}), and finally perform the inverse Fourier transform to get the anisotropic correlation function in configuration space.  In the computations, the voxel size is taken to be 5 $\MpcOh$, the same as \citet{Avila_etal2022}.  The parameters for $R_{\rm beam}$  and $k_{\rm fg} $  are indicated at the end of Sec.~\ref{sec:21cm_Pk}.

\subsection{Parameters and their inference  }
We describe the procedures for extracting the BAO scale from the correlation function measurement, following the standard template fitting method (e.g.,~\citet{Seo_etal2012, Xu_etal2012,  Chan_etal2018}).

The template and covariance are computed in the fiducial Planck cosmology.  To accommodate any discrepancy between the fiducial model and the true data cosmology, we introduce the parameters $\alpha_\perp $ and $\alpha_\parallel $ to rescale the wavenumber in the transverse and radial direction, respectively:
\begin{align}
  k'_\perp    & = \frac{ k_\perp }{ \alpha_\perp },          \\
  k'_\parallel & = \frac{ k_\parallel }{ \alpha_\parallel } .   
\end{align}
Physically, these scale dilation parameters rescale the BAO scale in the transverse and radial directions respectively
\begin{align}
\alpha_\perp  & = \frac{ D_{\rm M} r_{\rm s}^{\rm fid} }{ D_{\rm M}^{\rm fid}  r_{\rm s} },  \\
\alpha_\parallel  & = \frac{ D_{\rm H} r_{\rm s}^{\rm fid} }{ D_{\rm H}^{\rm fid}  r_{\rm s} }, 
\end{align}
where $ D_{\rm M} $ is the comoving angular diameter distance, $ D_{\rm H} $ denotes the Hubble distance $cH^{-1}(z)$, and $r_{\rm s} $ represents the sound horizon at the drag epoch.  The quantities with and without ``fid'' mean that they are evaluated in the fiducial and the data cosmology, respectively.

As in the standard method (e.g.,~\citet{Seo_etal2012, Chan_etal2018}), the full model in real space is given by
\beq
\xi_{\rm model} (r_\perp, r_\parallel) = B  \xi_{\rm template}( \alpha_\perp r_\perp, \alpha_\parallel r_\parallel )  + \sum_{i} a_i r^i,  
\eeq
where  $ \xi_{\rm template}$ is the correlation function template computed using Eq.~\eqref{eq:xi_rparallel}, \eqref{eq:multipole_correlation}, or \eqref{eq:muwedge_correlation}. We include an amplitude parameter $B$ to account for the uncertainty in $b_{\rm 21cm}$. Implicitly, $\xi_{\rm template }$  is a function of $R_{\rm beam}$ and $k_{\rm fg}$ as well. The broadband term $ \sum_{i} a_i r^i $ is a polynomial in $ r = \sqrt{ r_\perp^2 +  r_\parallel^2 } $, introduced to model the potential mismatch between the broadband shape of the template  and the actual measurement arising from, e.g., the inadequacy in nonlinearity modeling and variation in the broadband shape due to differences in cosmologies.  %In the fiducial setup, $i$ goes over 0, 1, and 2. 

We sample the posterior distribution using the MCMC sampler {\tt emcee} \citep{Emcee_Foreman-Mackey2013} with the fitting parameters
\begin{equation}
  \label{eq:general_template}
  \Theta=\left\{\alpha_{\perp}, \,  \alpha_{\parallel}, \,  R_{\rm beam}, \,  k_{\rm fg}, \, B,  \, a_i \right\} . 
\end{equation}
The data likelihood $L$ is taken to be Gaussian
\begin{align}
L  \propto  e^{- \chi^2 / 2 } , 
\end{align}
with  $\chi^2 $ given by
\begin{equation}
  \chi^2= ( \xi_{\rm data} - \xi_{\rm model} )_i C^{-1}_{\phantom{...} ij}   ( \xi_{\rm data} - \xi_{\rm model} )_j, 
\end{equation}
where  $ \xi_{\rm data}$ and $ \xi_{\rm model}$ denote the data and model data vectors, respectively, and  $ C $ is the covariance.

The prior distribution is assumed to be uniform within some interval. 
The prior ranges for both $\alpha_{\perp}$ and $\alpha_{\parallel}$ are $\left [0.7, 1.3\right ]$ in all cases except R38, for which we adopt $\left [0.5, 1.5\right ]$ for $ \alpha_\perp $. The prior range for $ R_{\rm beam } $ is within $ 10 \MpcOh $ from the true value.  The prior range for  $ k_{\rm fg} $ is  [0, 0.01]$\hOMpc$ ([0, 0.1]$\hOMpc$) for true $ k_{\rm fg} =  0.00364 \, \hOMpc $ ($0.0419  \, \hOMpc  $).   For each bin (radial bin, multipole bin, or wedge bin), we include a broadband polynomial with $i = 0$  and 1 in Eq.~\eqref{eq:general_template},  except for the monopole bin, for which  $i = 0$, 1 and 2.    We perform the fitting in the range  $52.5 \le r \le   147.5 \MpcOh$.

%%%%%%%%%%%%%%%%%%%%%%%%%%%%%%%%%%%%%%%%%%%%%%%%%%%%%%%%%%%%%%%%%%%%%%%%%%%%%%%%%%%%%%%%%%%%%%%%%%

\section{Results}
\label{sec:results} 

In this section, we first use the mock catalog to look for the optimal settings for the three kinds of correlation functions, respectively. We then apply these statistics to forecast the parameter constraint for the ongoing and forthcoming 21 cm IM BAO experiments.

\begin{table}[!tbp]
	\centering
        \caption{ Parameter constraints of radial correlation function $ \xi_{r \parallel} $  on $ \alpha_\parallel $, $R_{\rm beam}$, and  $k_{\rm fg} $ for $ r_{\perp \rm max} $ of 5, 10, 25, 50, and 100 $\MpcOh$. The results obtained with various observation conditions are compared.  The observation condition models are: $ R_{\rm beam} = 10, \, 38.45 \MpcOh  $ (denoted as R10, and R38), and  $k_{\rm fg }= 0, \, 0.00364, \, 0.0419  \hOMpc$ (denoted as k0, k1, and k2).  To guide the eyes, the constraint on  $\alpha_{\parallel}$ with the smallest error bar is highlighted in bold.   }
	\label{tab:radial_parameter_rperpmax}
        \resizebox{0.9\textwidth}{!}{\begin{minipage}{\textwidth}
	\begin{tabular}{l c c c c }
		\hline
		Config.  &  $r_{\bot \rm max} / \MpcOh$  & $\alpha_{\parallel}$ & $R_{\mathrm{beam}} /  \MpcOh $ & $k_{\mathrm{fg}}/ \hOMpc$ \\
		\hline
		R10k0          &  5            & $0.975^{+0.084}_{-0.128}$ & $12.49^{+5.25}_{-7.45}$ & -- \\
		
		          & 10             & $0.960^{+0.078}_{-0.123}$     & $11.83^{+5.68}_{-7.27}$ & -- \\
		
		          & 25             & $0.967^{+0.070}_{-0.121}$       & $11.50^{+5.95}_{-7.28}$ & -- \\
                
		          & 50           & \boldmath$0.982^{+0.052}_{-0.085}$ & $10.83^{+6.38}_{-7.24}$ & -- \\
		
		         &  100            & $0.974^{+0.062}_{-0.126}$ & $10.76^{+6.46}_{-7.19}$ & -- \\
		\hline                 
		R38k0          & 5              & $0.991^{+0.152}_{-0.170}$ & $40.65^{+5.44}_{-7.17}$ & -- \\
		
		          & 10             &  $0.986^{+0.151}_{-0.167}$        & $40.66^{+5.52}_{-7.25}$ & -- \\
		
		         & 25             &  $0.986^{+0.142}_{-0.165}$      & $40.51^{+5.57}_{-7.20}$ & -- \\
                
                         & 50             & $0.985^{+0.126}_{-0.160}$       & $40.50^{+5.56}_{-7.22}$ & -- \\      
		        &  100          & \boldmath $0.986^{+0.123}_{-0.160}$ & $40.15^{+5.79}_{-7.23}$ & -- \\       
		\hline                 
		R10k1          &  5             & $0.969^{+0.0803}_{-0.120}$ & $12.62^{+5.76}_{-7.36}$ & $0.005^{+0.004}_{-0.003}$ \\
		
		          &  10            &  $0.966^{+0.076}_{-0.116}$ & $11.95^{+5.55}_{-7.13}$ & $0.005^{+0.004}_{-0.004}$ \\
		
		          &  25            & $0.969^{+0.071}_{-0.112}$ & $11.78^{+5.76}_{-7.36}$ & $0.005^{+0.004}_{-0.003}$ \\
                
		          & 50            & \boldmath$0.983^{+0.051}_{-0.078}$ & $11.11^{+6.23}_{-7.24}$ & $0.005^{+0.004}_{-0.004}$ \\
		
		         &  100            & $0.976^{+0.063}_{-0.121}$ & $10.72^{+6.45}_{-7.14}$ & $0.005^{+0.004}_{-0.004}$ \\

                \hline  
		R38k1          & 5              & $0.986^{+0.159}_{-0.170}$ & $40.72^{+5.49}_{-7.14}$ & $0.005^{+0.004}_{-0.004}$ \\
		
		          & 10             & $0.991^{+0.151}_{-0.167}$ & $40.66^{+5.44}_{-7.10}$ & $0.005^{+0.004}_{-0.004}$ \\
		
		          &  25            &  $0.985^{+0.143}_{-0.162}$ & $40.39^{+5.54}_{-7.11}$ & $0.004^{+0.002}_{-0.002}$ \\
                
		          & 50             & $0.987^{+0.128}_{-0.160}$ & $40.58^{+5.54}_{-7.23}$ & $0.005^{+0.004}_{-0.004}$ \\ 
		
		         &  100            & \boldmath$0.989^{+0.121}_{-0.162}$ & $40.39^{+5.67}_{-7.21}$ & $0.005^{+0.004}_{-0.004}$ \\
                \hline   
           	R10k2          & 5              & $0.965^{+0.076}_{-0.130}$ & $9.27^{+3.37}_{-3.13}$ & $0.042^{+0.002}_{-0.002}$ \\
		
		          & 10             & $0.968^{+0.074}_{-0.133}$ & $9.03^{+3.62}_{-3.39}$ & $0.042^{+0.002}_{-0.002}$ \\

		          & 25             & $0.978^{+0.066}_{-0.116}$ & $8.21^{+4.99}_{-4.52}$ & $0.042^{+0.003}_{-0.002}$ \\
                
		          & 50             & \boldmath$0.988^{+0.048}_{-0.064}$ & $8.47^{+6.46}_{-5.57}$ & $0.043^{+0.003}_{-0.003}$ \\

		         &  100            & $0.981^{+0.058}_{-0.080}$ & $7.84^{+6.53}_{-5.21}$ & $0.043^{+0.003}_{-0.003}$ \\
		\hline  
		R38k2          & 5              & $0.987^{+0.156}_{-0.160}$ & $38.62^{+6.52}_{-6.56}$ & $0.045^{+0.009}_{-0.008}$ \\

		         & 10             & $0.993^{+0.153}_{-0.158}$ & $38.29^{+6.64}_{-6.29}$ & $0.045^{+0.008}_{-0.008}$ \\

		          & 25              & $0.991^{+0.139}_{-0.149}$ & $38.22^{+6.67}_{-6.42}$ & $0.045^{+0.008}_{-0.007}$ \\

		          & 50             & $0.986^{+0.117}_{-0.137}$ & $38.39^{+6.57}_{-6.64}$ & $0.045^{+0.007}_{-0.006}$ \\ 

		         &  100            & \boldmath$0.990^{+0.110}_{-0.118}$ & $38.42^{+6.65}_{-6.72}$ & $0.046^{+0.007}_{-0.006}$ \\
        \hline
	\end{tabular}
        \end{minipage}}
\end{table}

\begin{table}[!tbp]
	\centering
	\caption{Parameter constraints for different combinations of the multipoles across various observation configurations. We have shown the results for monopole only (0); monopole and quadrupole ($ 0 \oplus 2  $);  monopole, quadrupole and hexadecapole ($ 0 \oplus 2 \oplus 4  $); and monopole, quadrupole, hexadecapole and hexacontatetrapole ($ 0 \oplus 2 \oplus 4 \oplus 6 $).  The tightest constraints on $\alpha_\perp$ and $\alpha_\parallel$ are highlighted in bold.    }
        \label{tab:multipole_parameter_l024}
        \resizebox{0.8\textwidth}{!}{\begin{minipage}{\textwidth}
				\begin{tabular}{l l c c c c}
				\hline
				Config.  & Multipoles & $\alpha_\perp$ & $\alpha_\parallel$ & $R_{\mathrm{beam}} / \MpcOh $ & $k_{\mathrm{fg}}/ \hOMpc$ \\
				\hline
				R0k0 & 0 & $0.988^{+0.055}_{-0.055}$ & $0.986^{+0.123}_{-0.134}$ & -- & -- \\
				& $0\oplus2$ & $0.998^{+0.033}_{-0.032}$ & $0.996^{+0.054}_{-0.054}$ & -- & -- \\
				& $0\oplus2\oplus4$ &  \boldmath$0.987^{+0.032}_{-0.032}$ & $1.000^{+0.052}_{-0.053}$ & -- & -- \\
				& $0\oplus2\oplus4\oplus6$ & \boldmath$1.000^{+0.031}_{-0.033}$ & \boldmath$0.996^{+0.050}_{-0.053}$ & -- & -- \\
				\hline
				R10k0 & 0 & $0.997^{+0.115}_{-0.123}$ & $1.002^{+0.169}_{-0.156}$ & $10.040^{+5.750}_{-4.780}$ & -- \\
				& $0\oplus2$ & $1.000^{+0.112}_{-0.150}$ & $0.983^{+0.104}_{-0.095}$ & $11.960^{+3.350}_{-3.180}$ & -- \\
				& $0\oplus2\oplus4$ & $1.008^{+0.077}_{-0.115}$ & \boldmath$0.980^{+0.099}_{-0.076}$ & $12.140^{+2.480}_{-2.100}$ & -- \\
				& $0\oplus2\oplus4\oplus6$ & \boldmath$1.015^{+0.120}_{-0.082}$ & $0.971^{+0.075}_{-0.101}$ & $12.652^{+2.250}_{-2.038}$ & -- \\
				\hline
				R38k0 & 0 & $0.973^{+0.337}_{-0.352}$ & $0.959^{+0.192}_{-0.226}$ & $42.370^{+4.280}_{-5.840}$ & -- \\
				& $0\oplus2$ & $0.995^{+0.331}_{-0.332}$ & $0.963^{+0.172}_{-0.208}$ & $41.220^{+4.790}_{-5.920}$ & -- \\
				& $0\oplus2\oplus4$ & $1.096^{+0.356}_{-0.275}$ & $0.959^{+0.159}_{-0.157}$ & $44.050^{+2.720}_{-3.120}$ & -- \\
				& $0\oplus2\oplus4\oplus6$ & \boldmath$1.147^{+0.239}_{-0.322}$ & \boldmath$0.977^{+0.147}_{-0.164}$ & $43.962^{+2.485}_{-2.677}$ & -- \\
				\hline                                
				R0k1 & 0 & $1.006^{+0.067}_{-0.068}$ & $0.982^{+0.138}_{-0.149}$ & -- & $0.005^{+0.004}_{-0.004}$ \\
				& $0\oplus2$ & $0.998^{+0.037}_{-0.036}$ & $1.001^{+0.072}_{-0.070}$ & -- & $0.005^{+0.004}_{-0.003}$ \\
				& $0\oplus2\oplus4$ & $0.996^{+0.035}_{-0.035}$ & $1.008^{+0.067}_{-0.064}$ & -- & $0.003^{+0.003}_{-0.002}$ \\
				& $0\oplus2\oplus4\oplus6$ & \boldmath$0.997^{+0.034}_{-0.035}$ & \boldmath$1.007^{+0.063}_{-0.065}$ & -- & $0.040^{+0.003}_{-0.003}$ \\
				\hline
				R10k1 & 0 & $0.982^{+0.117}_{-0.141}$ & $1.009^{+0.170}_{-0.149}$ & $10.190^{+5.310}_{-4.480}$ & $0.005^{+0.003}_{-0.003}$ \\
				& $0\oplus2$ & $1.007^{+0.091}_{-0.121}$ & $0.980^{+0.101}_{-0.080}$ & $12.070^{+4.030}_{-2.540}$ & $0.006^{+0.005}_{-0.004}$ \\
				& $0\oplus2\oplus4$ & \boldmath$1.010^{+0.081}_{-0.114}$ & \boldmath$0.976^{+0.100}_{-0.080}$ & $12.050^{+2.980}_{-2.320}$ & $0.005^{+0.004}_{-0.003}$ \\
				& $0\oplus2\oplus4\oplus6$ & $1.019^{+0.128}_{-0.092}$ & $0.971^{+0.082}_{-0.106}$ & $12.871^{+2.255}_{-2.039}$ & $0.004^{+0.003}_{-0.003}$ \\
				\hline
				R38k1 & 0 & $1.086^{+0.349}_{-0.286}$ & $0.995^{+0.230}_{-0.201}$ & $43.000^{+3.820}_{-5.630}$ & $0.004^{+0.004}_{-0.003}$ \\
				& $0\oplus2$ & $0.992^{+0.345}_{-0.332}$ & $0.971^{+0.151}_{-0.179}$ & $46.650^{+5.780}_{-5.440}$ & $0.003^{+0.004}_{-0.002}$ \\
				& $0\oplus2\oplus4$ & $1.082^{+0.374}_{-0.287}$ & $0.982^{+0.161}_{-0.164}$ & $44.170^{+2.710}_{-3.450}$ & $0.003^{+0.004}_{-0.002}$ \\
				& $0\oplus2\oplus4\oplus6$ & \boldmath$1.101^{+0.277}_{-0.367}$ & \boldmath$0.995^{+0.159}_{-0.161}$ & $43.890^{+2.723}_{-2.910}$ & $0.004^{+0.009}_{-0.003}$ \\
				\hline                                
				R0k2 & 0 & $1.008^{+0.107}_{-0.100}$ & $0.989^{+0.130}_{-0.121}$ & -- & $0.041^{+0.010}_{-0.007}$ \\
				& $0\oplus2$ & $0.998^{+0.066}_{-0.059}$ & $0.996^{+0.061}_{-0.070}$ & -- & $0.040^{+0.003}_{-0.003}$ \\
				& $0\oplus2\oplus4$ & $0.999^{+0.048}_{-0.052}$ & $0.996^{+0.053}_{-0.053}$ & -- & $0.040^{+0.003}_{-0.003}$ \\
				& $0\oplus2\oplus4\oplus6$ & \boldmath$1.000^{+0.050}_{-0.049}$ & \boldmath$0.996^{+0.052}_{-0.051}$ & -- & $0.040^{+0.003}_{-0.003}$ \\
				\hline
				R10k2 & 0 & $1.000^{+0.175}_{-0.190}$ & $1.009^{+0.120}_{-0.127}$ & $9.990^{+5.340}_{-5.820}$ & $0.049^{+0.012}_{-0.009}$ \\
				& $0\oplus2$ & $1.002^{+0.159}_{-0.145}$ & $0.996^{+0.075}_{-0.090}$ & $9.640^{+3.110}_{-3.110}$ & $0.041^{+0.003}_{-0.003}$ \\
				& $0\oplus2\oplus4$ & $1.006^{+0.115}_{-0.132}$ & $0.986^{+0.081}_{-0.080}$ & $10.000^{+1.350}_{-1.110}$ & $0.041^{+0.004}_{-0.003}$ \\
				& $0\oplus2\oplus4\oplus6$ & \boldmath$1.015^{+0.132}_{-0.104}$ & \boldmath$0.983^{+0.071}_{-0.074}$ & $10.300^{+0.981}_{-1.141}$ & $0.041^{+0.004}_{-0.003}$ \\
				\hline
				R38k2 & 0 & $0.984^{+0.336}_{-0.353}$ & $0.972^{+0.195}_{-0.192}$ & $39.630^{+6.160}_{-7.130}$ & $0.071^{+0.020}_{-0.019}$ \\
				& $0\oplus2$ & $0.974^{+0.336}_{-0.353}$ & $0.950^{+0.146}_{-0.192}$ & $45.770^{+8.090}_{-9.680}$ & $0.044^{+0.007}_{-0.007}$ \\
				& $0\oplus2\oplus4$ & $0.961^{+0.326}_{-0.355}$ & $0.998^{+0.167}_{-0.166}$ & $43.870^{+3.210}_{-4.650}$ & $0.045^{+0.007}_{-0.006}$ \\
				& $0\oplus2\oplus4\oplus6$ & \boldmath$0.978^{+0.352}_{-0.327}$ & \boldmath$1.020^{+0.163}_{-0.159}$ & $44.239^{+2.895}_{-3.992}$ & $0.045^{+0.006}_{-0.006}$ \\
				\hline
			\end{tabular}
         \end{minipage}}
\end{table}

\begin{table}[!tbp]
	\centering
	\caption{ Parameter constraints for different $\mu$-wedge bin divisions.
          The bin division is expressed as  $ [\theta_{\rm min}, \theta_{\rm max}, \Delta \theta ] $ in degrees, where $\theta_{\rm min}$ and $ \theta_{\rm max}$ are the minimum and maximum angles of the $\mu$-wedge boundaries, and $\Delta \theta$ is the angular bin width of the $\mu$-wedges. The tightest constraints on $\alpha_\perp$ and $\alpha_\parallel$ are highlighted in bold.   }
	\label{tab:mu_wedge_comparison_table}
        %\resizebox{0.9\textwidth}{!}{\begin{minipage}{\textwidth}
        \resizebox{0.83\textwidth}{!}{\begin{minipage}{\textwidth}
	    \begin{tabular}{l l c c c c}
	      \hline
	      Config. &  bin division/$^\circ$   & $\alpha_\perp$ & $\alpha_\parallel$  &  $R_{\mathrm{beam}} /  \MpcOh $   &    $k_{\mathrm{fg}}/ \hOMpc$ \\
	      \hline
				R0k0    &  [0, 90, 90] & $1.017^{+0.051}_{-0.053}$ & $0.997^{+0.118}_{-0.116}$ & -- & -- \\
				&  [0, 90, 18] & \boldmath$0.980^{+0.026}_{-0.026}$ & \boldmath$1.000^{+0.041}_{-0.040}$ & -- & -- \\
				&  [0, 90, 9] & $0.981^{+0.027}_{-0.027}$ & $1.001^{+0.044}_{-0.041}$ & -- & -- \\
				&  [18, 72, 18] & $0.989^{+0.041}_{-0.042}$ & $0.995^{+0.051}_{-0.051}$ & -- & -- \\
				&  [18, 72, 9] & $0.990^{+0.039}_{-0.040}$ & $0.994^{+0.050}_{-0.050}$ & -- & -- \\
				&  [18, 72, 4.5] & $0.991^{+0.037}_{-0.038}$ & $0.994^{+0.047}_{-0.048}$ & -- & -- \\
				\hline
				R10k0    &  [0, 90, 90] & $1.018^{+0.121}_{-0.123}$ & $1.007^{+0.162}_{-0.176}$ & $12.00^{+3.35}_{-3.18}$ & -- \\
				&  [0, 90, 18] & $1.042^{+0.119}_{-0.083}$ & $0.954^{+0.067}_{-0.089}$ & $13.69^{+1.85}_{-1.87}$ & -- \\
				&  [0, 90, 9] & $1.042^{+0.106}_{-0.079}$ & $0.953^{+0.067}_{-0.085}$ & $13.62^{+1.48}_{-1.42}$ & -- \\
				&  [18, 72, 18] & $1.018^{+0.111}_{-0.078}$ & $0.969^{+0.066}_{-0.090}$ & $10.93^{+3.40}_{-2.85}$ & -- \\
				&  [18, 72, 9] & $1.012^{+0.105}_{-0.072}$ & $0.968^{+0.064}_{-0.085}$ & $10.56^{+3.28}_{-2.67}$ & -- \\
				&  [18, 72, 4.5] & \boldmath$1.024^{+0.094}_{-0.066}$ & \boldmath$0.970^{+0.058}_{-0.071}$ & $11.60^{+2.32}_{-2.18}$ & -- \\
				\hline
				R38k0    &  [0, 90, 90] & $1.106^{+0.270}_{-0.364}$ & $0.942^{+0.244}_{-0.188}$ & $41.20^{+4.97}_{-5.52}$ & -- \\
				&  [0, 90, 18] & $1.053^{+0.270}_{-0.278}$ & $1.056^{+0.117}_{-0.108}$ & $40.10^{+1.66}_{-1.82}$ & -- \\
				&  [0, 90, 9] & \boldmath$1.040^{+0.274}_{-0.272}$ & \boldmath$1.113^{+0.106}_{-0.102}$ & $38.69^{+1.31}_{-1.42}$ & -- \\
				&  [18, 72, 18] & $1.052^{+0.295}_{-0.338}$ & $0.940^{+0.160}_{-0.155}$ & $40.34^{+4.97}_{-6.15}$ & -- \\
				&  [18, 72, 9] & $1.033^{+0.300}_{-0.330}$ & $0.955^{+0.150}_{-0.154}$ & $38.79^{+5.34}_{-6.13}$ & -- \\
				&  [18, 72, 4.5] & $1.026^{+0.309}_{-0.335}$ & $0.966^{+0.152}_{-0.164}$ & $39.16^{+5.17}_{-6.04}$ & -- \\
				\hline
				R0k1    &  [0, 90, 90] & $1.013^{+0.104}_{-0.116}$ & $0.979^{+0.194}_{-0.177}$ & -- & $0.0053^{+0.0036}_{-0.0029}$ \\
				&  [0, 90, 18] & $0.975^{+0.027}_{-0.028}$ & $1.009^{+0.041}_{-0.038}$ & -- & $0.0029^{+0.0006}_{-0.0004}$ \\
				&  [0, 90, 9] & \boldmath$0.985^{+0.025}_{-0.025}$ & \boldmath$0.994^{+0.034}_{-0.032}$ & -- & $0.0040^{+0.0007}_{-0.0005}$ \\
				&  [18, 72, 18] & $0.992^{+0.043}_{-0.046}$ & $0.989^{+0.053}_{-0.053}$ & -- & $0.0040^{+0.0016}_{-0.0016}$ \\
				&  [18, 72, 9] & $0.994^{+0.044}_{-0.045}$ & $0.985^{+0.053}_{-0.055}$ & -- & $0.0044^{+0.0017}_{-0.0014}$ \\
				&  [18, 72, 4.5] & $0.994^{+0.040}_{-0.040}$ & $0.985^{+0.053}_{-0.052}$ & -- & $0.0042^{+0.0014}_{-0.0012}$ \\
				\hline
				R10k1    &  [0, 90, 90] & $1.001^{+0.135}_{-0.126}$ & $1.001^{+0.156}_{-0.171}$ & $10.32^{+5.02}_{-4.51}$ & $0.0033^{+0.0025}_{-0.0023}$ \\
				&  [0, 90, 18] & $1.023^{+0.118}_{-0.072}$ & $0.971^{+0.058}_{-0.078}$ & $12.29^{+1.50}_{-1.35}$ & $0.0046^{+0.0011}_{-0.0009}$ \\
				&  [0, 90, 9] & \boldmath$0.997^{+0.072}_{-0.058}$ & \boldmath$0.996^{+0.050}_{-0.051}$ & $11.41^{+0.91}_{-0.85}$ & $0.0043^{+0.0008}_{-0.0006}$ \\
				&  [18, 72, 18] & $1.020^{+0.121}_{-0.088}$ & $0.964^{+0.070}_{-0.094}$ & $10.69^{+3.02}_{-2.72}$ & $0.0044^{+0.0013}_{-0.0012}$ \\
				&  [18, 72, 9] & $1.003^{+0.096}_{-0.074}$ & $0.971^{+0.064}_{-0.076}$ & $10.10^{+2.25}_{-2.08}$ & $0.0044^{+0.0011}_{-0.0009}$ \\
				&  [18, 72, 4.5] & $1.010^{+0.084}_{-0.064}$ & $0.982^{+0.055}_{-0.065}$ & $10.50^{+1.76}_{-1.68}$ & $0.0039^{+0.0008}_{-0.0007}$ \\
				\hline
				R38k1    &  [0, 90, 90] & $1.097^{+0.278}_{-0.365}$ & $0.945^{+0.234}_{-0.189}$ & $42.86^{+3.86}_{-5.45}$ & $0.0031^{+0.0027}_{-0.0021}$ \\
				&  [0, 90, 18] & $0.989^{+0.307}_{-0.295}$ & $1.060^{+0.116}_{-0.108}$ & $40.31^{+1.79}_{-1.99}$ & $0.0030^{+0.0015}_{-0.0014}$ \\
				&  [0, 90, 9] & \boldmath$0.958^{+0.320}_{-0.277}$ & \boldmath$1.111^{+0.105}_{-0.100}$ & $38.81^{+1.41}_{-1.57}$ & $0.0031^{+0.0014}_{-0.0013}$ \\
                                &  [18, 72, 18] & $1.068^{+0.295}_{-0.358}$ & $0.940^{+0.155}_{-0.160}$ & $42.28^{+4.20}_{-5.75}$ & $0.0049^{+0.0032}_{-0.0023}$ \\
				&  [18, 72, 9] & $1.076^{+0.290}_{-0.345}$ & $0.942^{+0.161}_{-0.158}$ & $40.97^{+4.61}_{-5.49}$ & $0.0044^{+0.0028}_{-0.0020}$ \\
				&  [18, 72, 4.5] & $1.066^{+0.294}_{-0.337}$ & $0.946^{+0.159}_{-0.156}$ & $41.24^{+4.49}_{-5.31}$ & $0.0046^{+0.0027}_{-0.0020}$ \\
				\hline
				R0k2    &  [0, 90, 90] & $1.013^{+0.101}_{-0.099}$ & $1.010^{+0.117}_{-0.112}$ & -- & $0.0413^{+0.0102}_{-0.0066}$ \\
				&  [0, 90, 18] & $0.952^{+0.039}_{-0.040}$ & $1.028^{+0.051}_{-0.049}$ & -- & $0.0389^{+0.0008}_{-0.0008}$ \\
				&  [0, 90, 9] & \boldmath$1.034^{+0.035}_{-0.033}$ & \boldmath$0.949^{+0.042}_{-0.044}$ & -- & $0.0401^{+0.0004}_{-0.0004}$ \\
				&  [18, 72, 18] & $0.982^{+0.054}_{-0.058}$ & $1.010^{+0.071}_{-0.062}$ & -- & $0.0371^{+0.0062}_{-0.0055}$ \\
				&  [18, 72, 9] & $0.983^{+0.045}_{-0.047}$ & $1.006^{+0.059}_{-0.055}$ & -- & $0.0401^{+0.0051}_{-0.0050}$ \\
				&  [18, 72, 4.5] & $0.980^{+0.046}_{-0.048}$ & $1.005^{+0.059}_{-0.054}$ & -- & $0.0408^{+0.0050}_{-0.0050}$ \\
				\hline
				R10k2    &  [0, 90, 90] & $0.991^{+0.196}_{-0.164}$ & $1.015^{+0.125}_{-0.126}$ & $8.97^{+4.22}_{-5.20}$ & $0.0434^{+0.0068}_{-0.0055}$ \\
				&  [0, 90, 18] & $0.943^{+0.101}_{-0.084}$ & $1.016^{+0.073}_{-0.068}$ & $10.11^{+0.88}_{-0.85}$ & $0.0397^{+0.0015}_{-0.0014}$ \\
				&  [0, 90, 9] & \boldmath$0.935^{+0.092}_{-0.077}$ & \boldmath$1.029^{+0.070}_{-0.069}$ & $9.73^{+0.84}_{-0.84}$ & $0.0397^{+0.0015}_{-0.0014}$ \\
				&  [18, 72, 18] & $1.047^{+0.154}_{-0.134}$ & $0.967^{+0.091}_{-0.103}$ & $11.84^{+1.11}_{-1.20}$ & $0.0408^{+0.0031}_{-0.0031}$ \\
				&  [18, 72, 9] & $1.005^{+0.138}_{-0.109}$ & $0.985^{+0.081}_{-0.087}$ & $10.24^{+1.97}_{-1.87}$ & $0.0409^{+0.0035}_{-0.0037}$ \\
				&  [18, 72, 4.5] & $1.039^{+0.133}_{-0.108}$ & $0.956^{+0.070}_{-0.087}$ & $11.25^{+1.02}_{-1.03}$ & $0.0431^{+0.0027}_{-0.0026}$ \\
				\hline
				R38k2    &  [0, 90, 90] & $0.954^{+0.367}_{-0.325}$ & $0.992^{+0.180}_{-0.194}$ & $35.77^{+8.34}_{-10.34}$ & $0.0653^{+0.0128}_{-0.0130}$ \\
				&  [0, 90, 18] & $0.998^{+0.351}_{-0.335}$ & \boldmath$1.008^{+0.157}_{-0.160}$ & $40.53^{+2.26}_{-2.29}$ & $0.0403^{+0.0029}_{-0.0027}$ \\
				&  [0, 90, 9] & $0.912^{+0.391}_{-0.302}$ & $1.064^{+0.148}_{-0.167}$ & $38.66^{+1.55}_{-1.78}$ & $0.0409^{+0.0032}_{-0.0028}$ \\
				&  [18, 72, 18] & $0.990^{+0.340}_{-0.319}$ & $0.981^{+0.169}_{-0.168}$ & $37.96^{+3.44}_{-3.62}$ & $0.0436^{+0.0040}_{-0.0039}$ \\
				&  [18, 72, 9] & $1.007^{+0.331}_{-0.328}$ & $0.981^{+0.171}_{-0.171}$ & $38.26^{+3.26}_{-3.28}$ & $0.0431^{+0.0039}_{-0.0037}$ \\
                                 &  [18, 72, 4.5] & \boldmath$1.033^{+0.322}_{-0.336}$ & $0.993^{+0.166}_{-0.183}$ & $40.30^{+2.71}_{-2.74}$ & $0.0399^{+0.0036}_{-0.0033}$ \\
				\hline
			\end{tabular}
        \end{minipage}}
\end{table}

\subsection{ Finding the optimal parameter choices }
\label{sec:optimal_parameter_choices}

\begin{figure*}[!htb]
  \centering
  \includegraphics[width=0.97\linewidth]{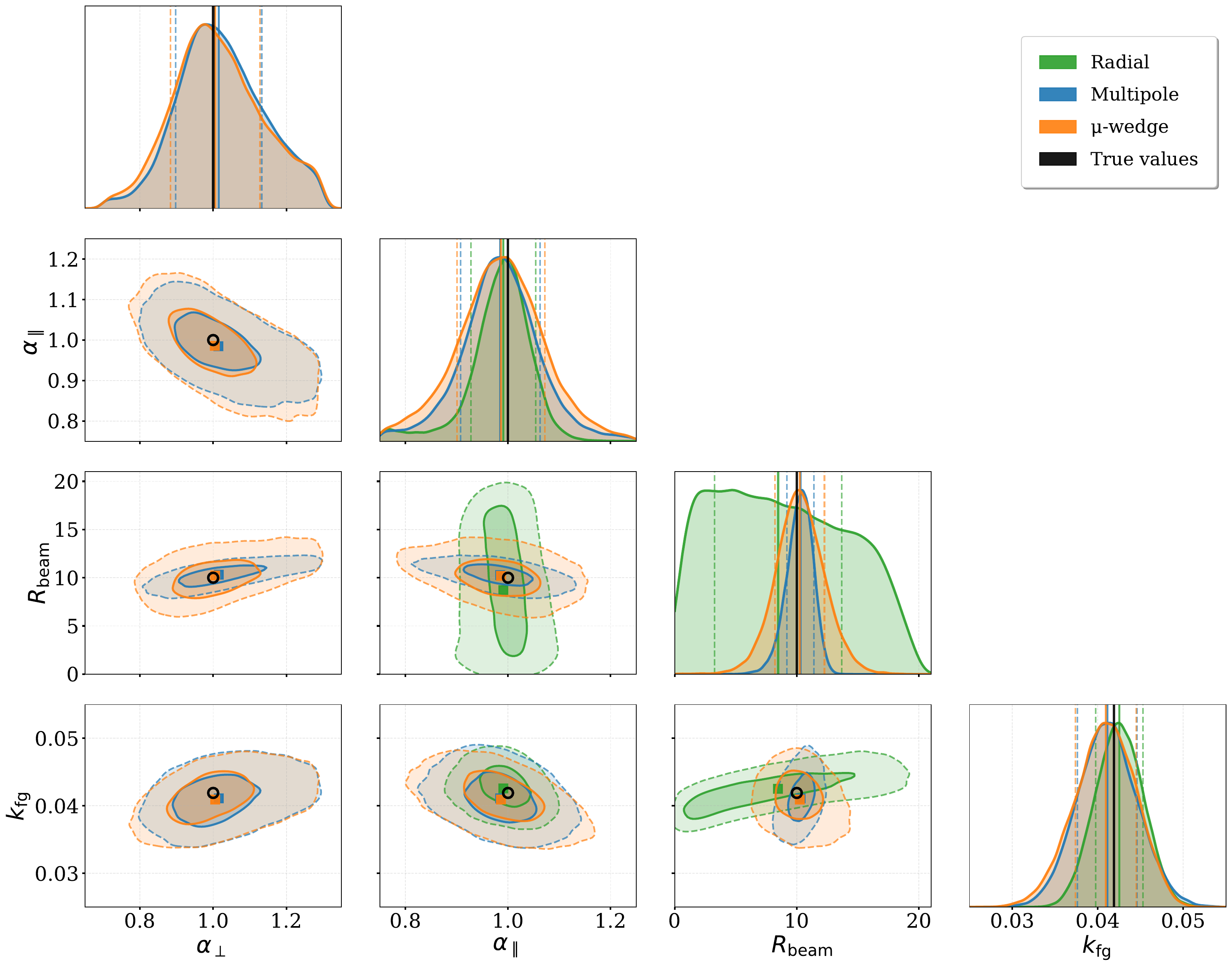}
  \caption{  Corner plots for  the parameters $\alpha_\perp$, $\alpha_\parallel$, $R_{\rm beam } $, and  $k_{\rm fg}$.  The posterior distributions for the radial correlation function (green), multipole correlation function (blue), and $\mu$-wedge correlation function (orange) are shown. The true values are highlighted in black. The inner and outer contour curves represent the 1-$\sigma$ and 2-$\sigma$ distribution boundaries, respectively, and the maximum of the distribution is marked by a square marker. In the 1D marginal distribution, we show the median (solid) and the 16th and 84th percentiles (dashed). }
 \label{fig:BAOfit_cornerplots}
\end{figure*}

In Table \ref{tab:radial_parameter_rperpmax}, we check the parameter constraint from the radial correlation function $ \xi_{r \parallel} $ with  $r_{\bot \rm max} =5, \, 10, \, 25, \,  50$, and 100  $\MpcOh$, respectively. Recall that we consider the observation conditions: $ R_{\rm beam} = 0, \, 10, \, 38.45 \MpcOh  $ (denoted as R0, R10, and R38, respectively in the tables), and  $k_{\rm fg }= 0, \, 0.00364, \, 0.0419  \hOMpc$ (denoted as k0, k1, and k2, respectively).    We show the best fits and their associated 1-$\sigma$ error bars. The best fit is estimated by the median of the posterior distribution, and the 1-$\sigma$ error bars by the 16th and 84th percentiles, respectively.  For easy inspection, the tightest constraint on $\alpha_\parallel $ is highlighted in bold.  Because finite $R_{\rm beam} $ is required to suppress the transverse BAO modes in $ \xi_{r \parallel} $, we only show $ R_{\rm beam} =  10 $ and  $38.45 \MpcOh $ results. For the $k_{\rm fg }= 0  \hOMpc$ case, we do not consider this parameter in the model.  Overall, the parameter constraint shows weak dependence on  $r_{\bot \rm max} $.  The $\alpha_\parallel $ constraint tightens  mildly with increasing  $r_{\bot \rm max} $, and it saturates for large  $r_{\bot \rm max} $.  We find that the tightest constraints come from either $r_{\bot \rm max} =50 $  or 100 $\MpcOh$, with no significant differences between them.  For  $R_{\rm beam } $ and $k_{\rm fg} $, we also do not find clear preference for  $r_{\bot \rm max} $ values.  Therefore, we take $r_{\bot \rm max} =50 \MpcOh $ in the rest of the analysis.

Table \ref{tab:multipole_parameter_l024} investigates the constraining power of the multipoles.  We have contrasted the results from various combinations of the  monopole ($l=0$), quadrupole ($l= 2 $), hexadecapole ($l= 4 $), and hexacontatetrapole ($l=6$).   For the isotropic case, we find that the introduction of the quadrupole significantly tightens the parameter constraint, while the inclusion of the higher multipoles (hexadecapole and hexacontatetrapole) does not give further substantial improvements.   The constraint on  $\alpha_\perp$ is stronger than the $\alpha_\parallel $ constraint because there are more perpendicular modes available.    Note that in the isotropic case, even though all the information is in the monopole, the addition of the higher-order multipoles does lead to higher information contents because of the confirmation of the vanishing of the higher-order multipoles. To compare the monopole-only case with monopole plus higher-order multipoles, we can think of the covariance of the higher-order multipoles in the monopole-only case as infinitely large, so that they do not contribute.  For the anisotropic cases due to $R_{\rm beam } $ or $ k_{\rm fg} $, $ 0 \oplus 2 $ and $ 0 \oplus 2 \oplus 4 $ both lead to successively tighter constraints in general, although  $ 0 \oplus 2 $ sometimes gives insignificantly tighter constraint than $ 0 \oplus 2 \oplus 4 $.  Inclusion of $l=6 $ offers even less significant improvement, signaling that $ 0 \oplus 2 \oplus 4 $ is already close to saturating the constraint.  We choose the combination $  0 \oplus 2 \oplus 4 \oplus 6 $  as the fiducial setup in the rest of the analysis.

%This trends are most clear on  $\alpha_\perp$,  $\alpha_\parallel$, and $R_\mathrm{beam} $, while there is not much difference between the $ 0 \oplus 2 $  and   $ 0 \oplus 2 \oplus 4  $ cases. 

We now turn to study the impact of the $\mu$ bin division on the  $\mu$-wedge correlation function constraint in Table \ref{tab:mu_wedge_comparison_table}.   Because the grid points of the  2D correlation function are uniformly distributed in the angular space, we choose to divide the wedge bins into equal angular bins.      We divide the angular range  $ [\theta_{\rm min}, \theta_{\rm max} ]$ into a number of bins, each with an equal angular width of $ \Delta \theta  $, and concisely express it as  $ [\theta_{\rm min}, \theta_{\rm max}, \Delta \theta ]$.

Let us first discuss the isotropic case. The [0, 90, 90]$^\circ$  case is equivalent to the monopole result in Table \ref{tab:multipole_parameter_l024}.  Indeed, the results are similar to the monopole ones, with the differences primarily caused by the number of broadband terms used.  We then divide the angular range into five bins ([0, 90, 18]$^\circ$) and ten bins ([0, 90, 9]$^\circ$), and we find that increase of resolution does not lead to tighter constraints. To contrast with the anisotropic cases, we also show the cases [18, 72, $\Delta \theta$]$^\circ$. With the modes at both ends cut out, the constraints are weaker than the full range case. We also show the highest resolution case, [18, 72, 4.5]$^\circ$, whose constraint is very similar to  [18, 72, 9]$^\circ$, showing that the constraints are already saturated at the resolution of $ \Delta \theta = 9 ^\circ$.

As we noted, for the anisotropic cases, there are artifacts appearing on the $x$ and $y$ axes in Fig.~\ref{fig:xi2D_model} due to the beam  and foreground removal effects, respectively. For the full range cases, especially [0, 90, 9]$^\circ$, the $\mu$-wedge correlation function often gives tighter constraint when the resolution increases. However, we also note that occasionally the bias is substantial in the sense that the expected values lie outside the 1-$\sigma$ interval. For example, in R38k0 and R0k2, [0, 90, 9]$^\circ$ gives the tightest but mildly biased constraint.  This demonstrates that these artifacts are difficult to model well and may cause the fit to be unstable.  The $\mu$-wedge correlation function allows for the possibility of removing the artifacts on the axes. By inspection of Fig.~\ref{fig:xi2D_model}, we choose to remove the intervals [0, 18]$^\circ$ and [72, 90]$ ^\circ $.  Excision of these intervals, on one hand, reduces the data size and hence enlarges the error bar,  while it gives rise to more stable fit on the other hand.  Overall, with the removal of the artifact regions, we find that the results are always within 1-$\sigma$ of the expected value. There are no significant improvements when the bin division is refined to  [18, 72, 4.5]$^\circ$. With these considerations in mind, \change{we will keep both bin divisions [0, 90, 9]$^\circ$ and [18, 72, 9]$^\circ$ for comparison in the forecast part, since the former gives a tight parameter constraint while the latter is robust to artifacts.   }

%In fact, finite grid size  of the $ \xi_{\rm 21cm }$ prevents the usage of too small wedge bin width because of the mismatch between the rectangular grid and wedge grid.  The discrete grid points can cause systematic fluctuations in the estimation of the bin values and in the worst case, there could be no grid points inside the bin.

We find that the configuration [0, 90, 9]$^\circ$ tends to give slightly tighter constraints than the multipole combination $ 0 \oplus 2 \oplus 4 \oplus 6$, while the fiducial setting [18, 72, 9]$^\circ$ yields compatible constraints as the multipole. This suggests that the lower multipoles are also insensitive to the artifacts.  This is so because the lower multipoles retain the ``wide-angle'' fluctuations and the impact of the artifacts is suppressed as they are relatively localized.

In Fig.~\ref{fig:BAOfit_cornerplots} we show the best-fit contour distributions for the parameters $\alpha_\perp$, $\alpha_\parallel$,  $R_{\rm beam} $, and  $ k_{\rm fg } $ obtained with their fiducial settings. \change{ This example uses $ R_{\rm beam } = 10 \MpcOh $ and $ k_{\rm fg } = 0.0419 \hOMpc$ (R10k2). We find that  the constraints by the multipole and $\mu$-wedge ([18,72,9]$^\circ$ configuration) are rather similar.  In contrast, the [0,90,9]$^\circ$  configuration would give a tighter constraint than the multipole, but it would also sometimes result in a bias at the 1$\sigma$ level.}  Besides the mild degeneracy between  $\alpha_\perp $ and   $\alpha_\parallel $,  there is no strong degeneracy between the parameters,  in particular, between $R_{\rm beam} $ and   $ k_{\rm fg} $.  While $R_{\rm beam} $  is well constrained by the multipoles or wedges, it is hardly constrained by the radial correlation function.

\begin{figure}[!tb]
  \centering
  \includegraphics[width=\linewidth]{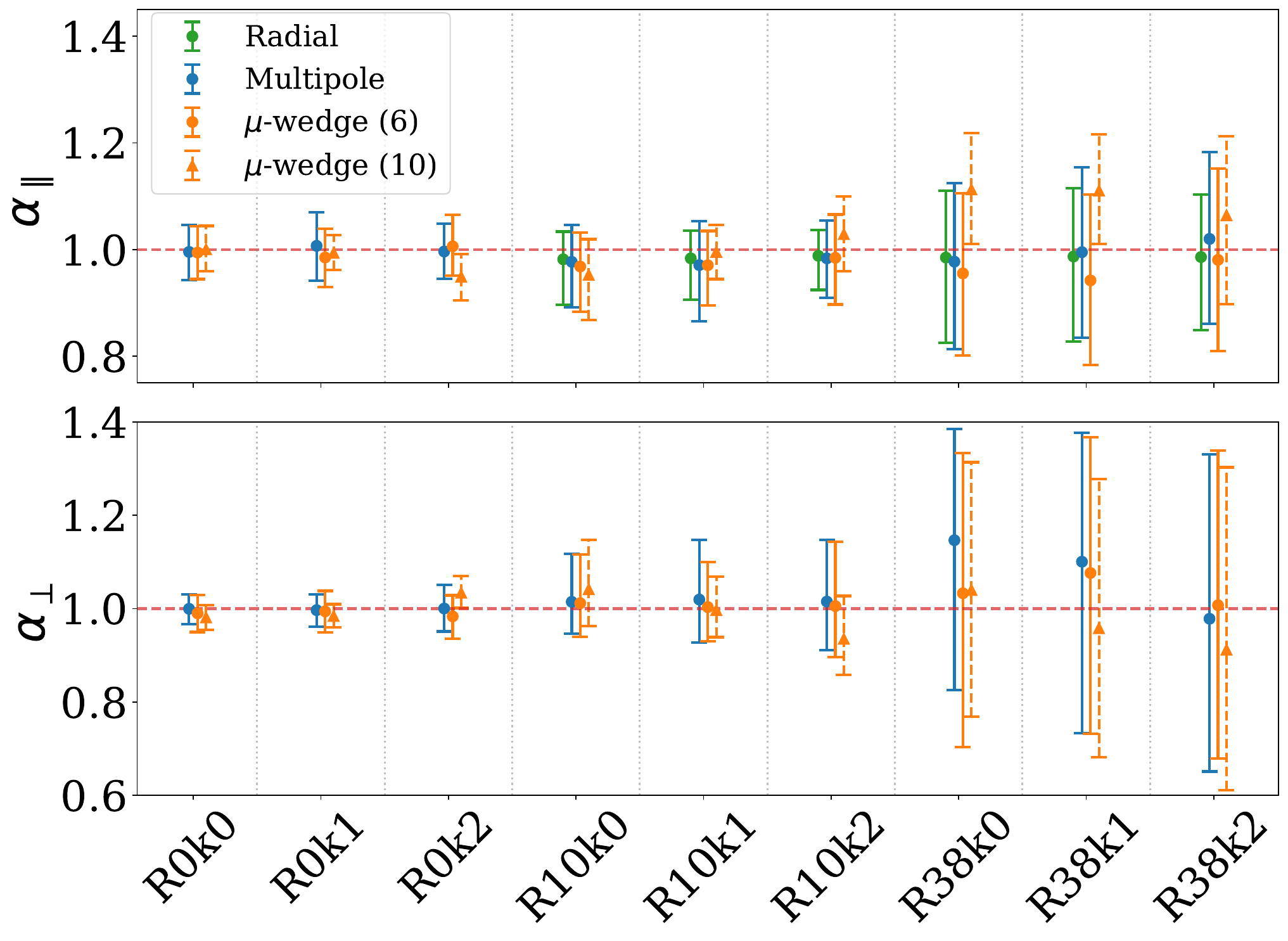}
  \caption{ Visual representation of the  best-fit $\alpha_\parallel$ and  $\alpha_\perp$  under various  observation conditions. The results from the radial correlation function (green), multipole correlation function (blue), and $\mu$-wedge correlation function (orange) are compared. \change{ For the $\mu$-wedge case, we include the six-wedge ([18,72,9]$^\circ$, circle markers with solid error bars) and 10-wedge ([0,90,9]$^\circ$, triangle markers with dashed error bars) cases. } }
  \label{fig:BAOfit_results}
\end{figure}

Fig.~\ref{fig:BAOfit_results}  visually contrasts the optimal setting results from these three statistics.  The best-fit results for $\alpha_\perp$ and  $\alpha_\parallel$  are shown across various observation conditions. Generally, the constraints become weaker when  $R_{\rm beam }$ increases, while the trend is not apparent for the $ k_{\rm fg} $ values considered. For both  $\alpha_\parallel $ and  $\alpha_\perp $, the multipoles and $\mu$-wedges show very comparable constraining power.  $ \xi_{r \parallel}$ can deliver an effective constraint on  $\alpha_\parallel $  for  $ R \ge 10 \MpcOh $ cases, and in fact, it outperforms the others by a small margin for the k2 cases. \change{ Although the 10-$\mu$ wedge case is generally more constraining than the six-wedge case, it is also less robust.  }

\begin{table*}[!tbp]
	\centering
 \caption{Constraints on $\alpha_\perp$ and $\alpha_\parallel$, $R_{\mathrm{beam}}$, and $k_{\mathrm{fg}}$  for different beam  and foreground removal models. The underlying data model is R10k2, generated with the fiducial Gaussian beam  and Gaussian foreground removal model, while the fitting beam model and foreground removal model are shown in the second column.   }
  \label{tab:beam_foreground_results}        
	\resizebox{0.90\textwidth}{!}{\begin{minipage}{\textwidth}
			\begin{tabular}{l c c c c c}
				\hline
    statistics   &beam + foreground removal& $\alpha_\perp$ & $\alpha_\parallel$   &  $R_{\mathrm{beam}} /  h^{-1}\mathrm{Mpc} $   &    $k_{\mathrm{fg}}/ h\,\mathrm{Mpc}^{-1}$   \\
    \hline
     Radial &Gaussian + Gaussian  &--&$0.988^{+0.048}_{-0.064}$ & $8.47^{+6.46}_{-5.57}$ & $0.043^{+0.003}_{-0.003}$ \\
    &Gaussian + $k_{\parallel,\mathrm{min}}$  &--&$0.927^{+0.078}_{-0.101}$ & $6.39^{+7.02}_{-4.60}$ & $0.033^{+0.003}_{-0.002}$ \\
    &Gaussian + cosine  &--&$0.988^{+0.049}_{-0.072}$ & $8.63^{+6.14}_{-5.33}$ & $0.048^{+0.003}_{-0.003}$ \\
    &Gaussian + jinc  &--&$0.960^{+0.062}_{-1.25}$ & $11.30^{+5.63}_{-5.54}$ & $0.047^{+0.004}_{-0.004}$ \\
    &Jinc + Gaussian & -- & $0.981^{+0.056}_{-0.084}$ & $9.62^{+6.51}_{-5.72}$ & $0.043^{+0.003}_{-0.003}$ \\
     &Cosine + Gaussian & --&$0.988^{+0.048}_{-0.061}$ & $7.99^{+6.31}_{-5.44}$ & $0.043^{+0.003}_{-0.003}$ \\
     \hline
    Multipoles &Gaussian + Gaussian& $1.015^{+0.132}_{-0.104}$ & $0.983^{+0.071}_{-0.074}$ & $10.30^{+0.98}_{-1.14}$ & $0.041^{+0.004}_{-0.003}$ \\
    &Gaussian + $k_{\parallel,\mathrm{min}}$  & $1.143^{+0.110}_{-0.149}$ & $0.927^{+0.090}_{-0.094}$ & $9.85^{+1.09}_{-1.32}$ & $0.034^{+0.004}_{-0.003}$ \\
    &Gaussian + cosine  & $0.990^{+0.144}_{-0.118}$ & $0.981^{+0.080}_{-0.084}$ & $10.52^{+1.11}_{-1.26}$ & $0.046^{+0.004}_{-0.004}$ \\
    &Gaussian + jinc  & $0.960^{+0.100}_{-0.115}$ & $0.982^{+0.080}_{-0.084}$ & $11.19^{+1.28}_{-1.38}$ & $0.041^{+0.005}_{-0.005}$ \\
    &Jinc + Gaussian & $0.968^{+0.120}_{-0.101}$ & $1.001^{+0.082}_{-0.077}$ & $10.21^{+1.13}_{-1.32}$ & $0.042^{+0.004}_{-0.004}$ \\
    &Cosine + Gaussian & $0.991^{+0.092}_{-0.080}$ & $1.001^{+0.066}_{-0.064}$ & $9.12^{+1.00}_{-1.13}$ & $0.041^{+0.003}_{-0.003}$ \\
    \hline
    $\mu$-wedge & Gaussian + Gaussian &$1.005^{+0.138}_{-0.109}$ & $0.985^{+0.081}_{-0.087}$ & $10.24^{+1.97}_{-1.87}$ & $0.041^{+0.004}_{-0.004}$ \\				
    &Gaussian + $k_{\parallel,\mathrm{min}}$  & $1.100^{+0.138}_{-0.168}$ & $0.964^{+0.101}_{-0.098}$ & $10.50^{+1.17}_{-1.34}$ & $0.043^{+0.003}_{-0.003}$ \\
    &Gaussian + cosine  & $1.040^{+0.146}_{-0.124}$ & $0.958^{+0.083}_{-0.104}$ & $11.10^{+1.07}_{-1.17}$ & $0.046^{+0.003}_{-0.003}$ \\
    &Gaussian + jinc  & $1.010^{+0.147}_{-0.120}$ & $0.955^{+0.104}_{-0.157}$ & $11.50^{+1.15}_{-1.19}$ & $0.037^{+0.003}_{-0.003}$ \\
    &Jinc + Gaussian & $0.996^{+0.154}_{-0.127}$ & $0.981^{+0.091}_{-0.105}$ & $11.00^{+1.40}_{-1.43}$ & $0.043^{+0.003}_{-0.003}$ \\
    &Cosine + Gaussian & $1.020^{+0.119}_{-0.093}$ & $0.970^{+0.066}_{-0.076}$ & $10.30^{+1.05}_{-1.09}$ & $0.043^{+0.003}_{-0.003}$ \\
    \hline
  \end{tabular}
  \end{minipage}}
\end{table*}

\subsection{ Impact of mismodeling of the beam and foreground removal models}
\label{sec:Impact_Mismodeling}

\change{ To assess the impact of mismodeling the beam shape or foreground removal model on the estimation of the parameters $\alpha_\perp$ and $\alpha_\parallel$, we use different models for the data and the fitting model.   The data are generated with the fiducial Gaussian beam and Gaussian foreground removal, while in the model, the beam is modeled either as a Gaussian, a jinc, or a cosine (see \citet{Matshawule_etal2021} for the exact expressions) and, in a similar spirit, we adopt a Gaussian, a jinc, a cosine, or a sharp $k_{\parallel \rm min}$ window for the foreground removal model.  We fit these models to the generated data, and the effective beam size $R_{\mathrm{beam}}$ and the foreground cut scale $k_{\mathrm{fg}}$ are allowed to vary.  }

\change{  Table~\ref{tab:beam_foreground_results} summarizes the results for the three statistics.   We consider  the configuration R10k2 here.   The main focus is on the stability of these  models relative to the fiducial Gaussian models.  For the Gaussian beam with various foreground removal models, we find that the worst case is the aggressive $k_{\parallel,\mathrm{min}}$ cut, which results in $ \alpha_\perp $ and $\alpha_\parallel$ being biased high and low, respectively, albeit still within the inflated 1$\sigma$ bound.   For the cosine and jinc cuts, there are also biases in the fit, although they are less severe. When we use a jinc or cosine beam with Gaussian foreground removal, the impact tends to be milder, and we observe only small biases relative to the fiducial cases. Thus, this exercise suggests that the foreground modeling has a bigger impact on the BAO measurement. As in the fiducial case results shown in Fig.~\ref{fig:BAOfit_cornerplots}, we also do not find significant correlation between $R_{\mathrm{beam}}$ and  $k_{\mathrm{fg}}$ in these cases.    }

% First table: Instrument parameters
\begin{table*}[!tbp]
  \centering
\caption{ Key parameters of the major 21 cm IM  surveys and their instruments used for BAO constraint forecasts.     }
\label{tab:surveys_parameters}
\resizebox{\textwidth}{!}{
 \begin{minipage}{1.15\textwidth}
\begin{tabular}{c|cccccccc ccc}
  %\toprule
  \hline
Experiment & $\bar{z}$ & FWHM /deg & $r(z) /\MpcOh $ & $l_z /\MpcOh $ & $f_{\text{sky}}$ & $T_{\text{sys}}$ /\,K & $t_{\text{tot}} $ /\,h &  $ N_{\rm dish} \times  N_{\text{feed}}$    &   $R_{\text{beam}}/ \MpcOh $ & $ k_{\rm fg}/ \hOMpc $  & $ P_{\rm N } /  ({\rm mK})^2 (\MpcOh)^{-3}    $   \\ 
%\midrule
\hline
%SKA1 & 1.32 & 1.85 & 2793 & 300 & 0.12 & 26 & 10000 & $197 \times 1$ \\
BINGO   & 0.29    & 0.38 & 800  & 850              & 0.15  & 70 & 17520 & $1 \times 50$   & 2.27  &  0.015 & 2.9   \\
MeerKAT & 0.39  & 1.25 & 1000 & 342             & 0.1   & 19 & 4000  & $64 \times 1$   & 9.85  &  0.037    & 0.9  \\
SKA-mid & 1.32    & 1.85 & 2793 & 3450            & 0.48  & 26 & 10000 & $197 \times 1$  & 38.45 &  0.0036 & 12.2 \\
%\bottomrule
\hline
\end{tabular}
\end{minipage}}
\end{table*}

\begin{table*}[!tbp]
	\centering
	\caption{ Forecasts for the BAO recovery from the major 21 cm experiments. We have shown the forecast results from three statistics using the optimal settings. Both the results without and with RSD are compared.   }
	\label{tab:21cmexperiments_parameterconstraints}
	\resizebox{0.90\textwidth}{!}{\begin{minipage}{\textwidth}
			\begin{tabular}{l l c c c c}
				\hline
	         Experiment     & statistics   & $\alpha_\perp$ & $\alpha_\parallel$   &  $R_{\mathrm{beam}} /  \MpcOh $   &    $k_{\mathrm{fg}}/ \hOMpc$   \\
				\hline
        	BINGO           & Radial  w/o RSD & -- & $0.976^{+0.060}_{-0.083}$ & $6.48^{+3.98}_{-4.27}$ & $0.0115^{+0.0046}_{-0.0083}$ \\
				& Multipole  w/o RSD & $0.998^{+0.045}_{-0.043}$ & $1.001^{+0.063}_{-0.063}$ &  $2.32^{+1.43}_{-1.58}$ & $0.0151^{+0.0025}_{-0.0032}$ \\
                & $\mu$-wedge (6)  w/o RSD &  $0.997^{+0.043}_{-0.043}$ & $ 0.998^{+0.056}_{-0.056}$ & $2.39^{+1.44}_{-1.60}$ & $0.0156^{+0.0023}_{-0.0020}$ \\
		        & $\mu$-wedge (10)  w/o RSD &  $\mathbf{ 0.999^{+0.030}_{-0.029}}$ & $ \mathbf{1.000^{+0.043}_{-0.043}}$ & $2.26^{+0.14}_{-0.14}$ & $0.0152^{+0.0004}_{-0.0004}$ \\
			 	& Radial & -- & $0.986^{+0.056}_{-0.084}$ & $6.35^{+4.13}_{-4.30}$ & $0.0121^{+0.0061}_{-0.0086}$ \\
                                 & Multipole  & $1.000^{+0.035}_{-0.038}$ & $1.007^{+0.054}_{-0.044}$ &  $2.24^{+1.36}_{-1.41}$ & $0.0155^{+0.0020}_{-0.0018}$ \\
			 	& $\mu$-wedge (6)  &  $1.001^{+0.039}_{-0.040}$ & $0.998^{+0.025}_{-0.027}$ & $2.37^{+1.48}_{-1.54}$ & $0.0153^{+0.0002}_{-0.0002}$ \\
                               & $\mu$-wedge (10)  &  $ \mathbf{1.000^{+0.030}_{-0.030} }$ & $\mathbf{ 0.998^{+0.019}_{-0.021}}$ & $2.27^{+0.16}_{-0.16}$ & $0.0152^{+0.0004}_{-0.0004}$ \\
				\hline
				MeerKAT   & Radial  w/o RSD & -- & $0.994^{+0.105}_{-0.119}$ & $9.42^{+6.23}_{-6.26}$ & $0.0366^{+0.0034}_{-0.0031}$ \\
				& Multipole  w/o RSD & $1.009^{+0.092}_{-0.082}$ & $0.996^{+0.096}_{-0.103}$ &   $9.86^{+0.64}_{-0.61}$ & $0.0370^{+0.0023}_{-0.0021}$ \\
                & $\mu$-wedge (6)  w/o RSD &  $1.007^{+0.083}_{-0.070}$ & $0.991^{+0.073}_{-0.080}$ & $9.90^{+0.47}_{-0.43}$ & $0.0368^{+0.0018}_{-0.0017}$ \\
				& $\mu$-wedge (10)  w/o RSD &  $\mathbf{1.007^{+0.060}_{-0.055}}$ & $\mathbf{0.991^{+0.058}_{-0.062}}$ & $9.91^{+0.29}_{-0.27}$ & $0.0368^{+0.0009}_{-0.0008}$ \\
				%	            & Multipole &$1.005^{+0.113}_{-0.098}$ & $0.985^{+0.104}_{-0.118}$ & $10.03^{+1.07}_{-1.17}$ & $0.0376^{+0.0033}_{-0.0030}$ \\
                                & Radial & -- & $1.020^{+0.103}_{-0.103}$ & $8.73^{+6.13}_{-5.93}$ & $0.0363^{+0.0040}_{-0.0041}$ \\
				& Multipole  & $1.003^{+0.079}_{-0.069}$ & $0.983^{+0.062}_{-0.064}$ &  $10.12^{+0.97}_{-0.97}$ & $0.0373^{+0.0026}_{-0.0023}$ \\
				& $\mu$-wedge (6)  &  $1.003^{+0.064}_{-0.063}$ & $0.990^{+0.034}_{-0.042}$ & $9.97^{+0.55}_{-0.53}$ & $0.0370^{+0.0002}_{-0.0002}$ \\
                                & $\mu$-wedge (10) &  $\mathbf{0.993^{+0.046}_{-0.047}}$ & $ \mathbf{0.991^{+0.021}_{-0.028}}$ & $9.95^{+0.30}_{-0.27}$ & $0.0367^{+0.0008}_{-0.0008}$ \\
				\hline
				SKA-mid   & Radial  w/o RSD  & -- & $\mathbf{1.001^{+0.006}_{-0.006}}$ & $38.12^{+4.62}_{-4.23}$ & $0.0042^{+0.0034}_{-0.0029}$ \\
				& Multipole  w/o RSD  & $1.001^{+0.043}_{-0.043}$ & $1.000^{+0.007}_{-0.007}$ &  $38.45^{+0.14}_{-0.14}$ & $0.0037^{+0.0003}_{-0.0003}$ \\
                & $\mu$-wedge (6)   w/o RSD  &  $1.010^{+0.050}_{-0.046}$ & $1.000^{+0.013}_{-0.013}$ & $38.50^{+0.56}_{-0.56}$ & $0.0037^{+0.0002}_{-0.0002}$ \\
                & $\mu$-wedge (10)   w/o RSD  &  $\mathbf  {1.001^{+0.035}_{-0.033}}$ & $1.000^{+0.010}_{-0.010}$ & $38.45^{+0.14}_{-0.14}$ & $0.0036^{+0.0002}_{-0.0002}$ \\
				& Radial & -- & $ \mathbf{1.001^{+0.004}_{-0.005}}$ & $38.02^{+1.78}_{-1.67}$ & $0.0051^{+0.0038}_{-0.0033}$ \\
				& Multipole  & $\mathbf{0.998^{+0.022}_{-0.021}}$ & $1.001^{+0.005}_{-0.005}$ &  $38.46^{+0.09}_{-0.09}$ & $0.0037^{+0.0005}_{-0.0005}$ \\			
				& $\mu$-wedge (6)  &  $1.001^{+0.041}_{-0.038}$ & $1.000^{+0.009}_{-0.009}$ & $38.50^{+0.50}_{-0.50}$ & $0.0037^{+0.0003}_{-0.0003}$ \\
                & $\mu$-wedge (10)  &  $1.000^{+0.026}_{-0.030}$ & $1.000^{+0.007}_{-0.007}$ & $38.50^{+0.112}_{-0.120}$ & $0.0037^{+0.0002}_{-0.0002}$ \\
				\hline
			\end{tabular}
	\end{minipage}}
\end{table*}

\begin{figure}[!tb]
  \centering
  \includegraphics[width=\linewidth]{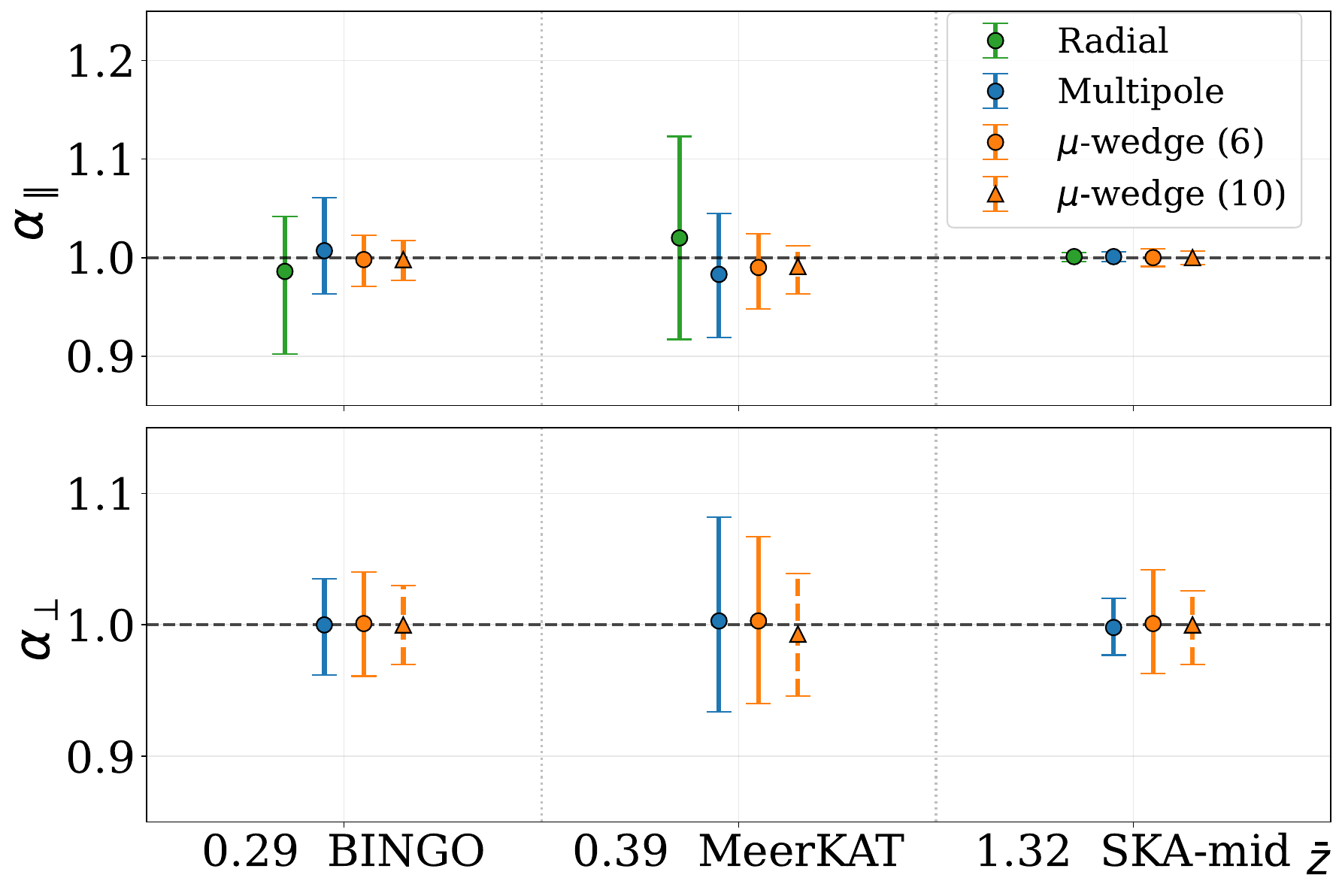}
  \caption{ Visual representation of the  best-fit  $\alpha_\parallel$ and $\alpha_\perp$  for BINGO, MeerKAT, and SKA-mid. On the $x$-axis, the mean redshifts of these surveys are labeled. The results from the radial correlation function (green), multipole correlation function (blue), and $\mu$-wedge correlation (orange) are compared. \change{    For the $\mu$-wedge case, we include the the six-wedge ([18,72,9]$^\circ$, circle markers with solid error bars) and 10-wedge ([0,90,9]$^\circ$, triangle markers with dashed error bars) cases.  }   }
  \label{fig:all_telescopes_BAO_fit}
\end{figure}

\subsection{Survey forecasts}

After verifying and optimizing the pipelines with the mocks, we now apply them to forecast the parameter constraints for the major ongoing and forthcoming 21 cm IM surveys.   They encompass  BINGO \citep{Abdalla_etal2022,Wuensche_etal2022}, MeerKAT \citep{Santos_etal2016}, SKA \citep{Braun_etal2015,Maartens_etal2015,Bacon_SKA2020}. The key parameters of these surveys and their instruments used for the forecasts are listed in Table \ref{tab:surveys_parameters}.  The pathfinding experiments BINGO and MeerKAT focus on the relatively low-redshift universe, while the mean redshift and volume of the SKA are larger than its predecessors. In particular, for SKA-mid, we consider the Wide Band 1 survey setting \citep{Bacon_SKA2020}, which is an IM survey covering 20,000 $\mathrm{deg}^2 $ in the redshift range of $0.35<z<3$.  The foreground parameter $ k_{\rm fg } $ is computed by taking $ N_\parallel =2 $ in Eq.~\eqref{eq:kfg_prescription}, the same as the prescription used in the first part. \change{ In this part, the results take RSD into account unless otherwise specified. }

To perform the forecast, we generate a data vector using the theoretical model on a grid with spacing of  1  $\MpcOh$ and then apply the fitting pipeline to this mock data vector to get the constraint. The forecasts on the parameters are shown in Table \ref{tab:21cmexperiments_parameterconstraints}. To highlight the constraints (with RSD) on $\alpha_\perp $ and $\alpha_\parallel$, we present them as a function of the mean redshifts of these surveys in Fig.~\ref{fig:all_telescopes_BAO_fit} as well.  \change{  Here we have contrasted the results obtained without and with RSD in Table  \ref{tab:21cmexperiments_parameterconstraints}. We find that the constraints tend to tighten when RSD is included, especially for $ \alpha_\parallel $, thanks to the enhanced clustering in the radial direction in redshift space.   Our results are discrepant with \citet{KennedyBull_2021}, who instead reported that RSD weakens the constraint on  $\alpha_\perp $ strongly and  on $\alpha_\parallel$ mildly.    }

\change{ We find that for BINGO, the tightest constraint on $\alpha_\parallel$  is achieved by the $\mu$-wedge, while multipole and $\mu$-wedge give similar constraining results on $\alpha_\perp$.
For MeerKAT, the tightest constraints on both $\alpha_\parallel$ and $\alpha_\perp$ are attained by the $\mu$-wedge statistic.  For SKA-mid, multipole and wedge give comparable constraints on  $\alpha_\perp$, with multipole being marginally tighter, while the radial correlation function is the most constraining on $\alpha_\parallel$. }

\change{ It turns out that the optimal statistics are only slightly altered when we move from  real space to redshift space.   This may suggest that our optimized parameter choices obtained in Sec.~\ref{sec:optimal_parameter_choices} are only slightly less optimal in the presence of RSD.
  Moreover, Table \ref{tab:21cmexperiments_parameterconstraints} shows that the 10-wedge setting ([0,90,9]$^\circ$) is more constraining than the six-wedge one.  However, in this study, because we fit to the mean model without noise, it is not clear if noise can cause instability in the fit. We have performed checks by adding noise to the mean model according to the covariance. We find that for these experiments, the 10-wedge case is {\it not} more unstable than the six-wedge setting in general. More specifically, for BINGO, the 10-wedge case is only slightly less stable. The case for MeerKAT is the opposite; the six-wedge case not only has larger error bars but is also less stable. For SKA-mid, the 10-wedge case gives a marginally tighter bound without an increase in instability. Thus, for these experiments, the full-wedge setting is recommended. }

%For the low redshift experiments BINGO and MeerKAT, the $R_{\rm beam }$  size is small or moderate. We find that the $\mu$-wedge is the most effective in constraining  $\alpha_\perp $ or $\alpha_\parallel$.  For SKA-mid, which has a larger $ R_{\rm beam}$ and smaller $k_{\rm fg}$, while $\mu$-wedge still attains the strongest constraint on $ \alpha_\perp $,  $\xi_{r \parallel} $ is the most constraining one for $\alpha_\parallel $.

We wish to compare our results with others', and so we consider the total $(S/N)^2 $ defined as
\beq
\bigg( \frac{S}{N} \bigg)^2 = \sum_i \bigg(  \frac{ \alpha_i }{ \sigma_{\alpha,i} } \bigg)^2 \approx \sum_i \bigg(  \frac{ 1 }{ \sigma_{\alpha,i} }\bigg)^2,   
\eeq
where the summation runs over all the tomographic bins, and $ \alpha_i$ and $ \sigma_{\alpha,i} $ denote the best fit and its error bar in the $i$th bin.  This is especially useful if others' forecast is conducted in multiple tomographic bins while ours is computed at the mean redshift only.

%For BINGO, while the constraint on $ \alpha_\parallel $ by the $\mu$-wedge is tighter than the multipole by a factor of two only or so, the constraint on $ \alpha_\perp $ is tighter by more than an order of magnitude. The strong  $ \alpha_\perp $ constraint, especially by the $\mu$-wedge, can be attributed to the small $ R_{\rm beam } $ in  BINGO. For the same reason, the radial correlation function is not particularly useful in this case.  

\citet{Novaes_etal2022} performed a BAO constraint forecast for BINGO using only the transverse information. In that study, the whole redshift range is divided into 30 fine tomographic bins, and the angular correlation functions are measured.  Using their fit to the mean angular correlation function (Table 2 of \citet{Novaes_etal2022}), we get $(S/N)^2 = 178$ on $ \alpha_\perp$\footnote{We note that \citet{Novaes_etal2022} also performed the angular power spectrum analysis, and it yielded instead a much higher $ (S/N)^2 $, 1700. }.  Our multipole and $\mu$-wedge analysis result in $ (S/N)^2 $ of 516 and 1150, respectively. Moreover, we show that appreciable constraint on $\alpha_\parallel $ is also achievable by the multipoles and $\mu$-wedges, with  $( S/N )^2 $ of 252 and 540, respectively.  Our results thus suggest that the 3D analysis is more effective than the multibin tomographic one in both  $\alpha_\perp $  and $\alpha_\parallel $ constraints.

We now turn to the MeerKAT survey. We first note that its $ R_{\rm beam }$ and $k_{\rm fg} $ values are  close to the R10k2 case, but there are two key differences, which are competing in BAO constraints. First, the MeerKAT redshift is lower, and hence the cosmological BAO signals are weaker. Second, the value of the noise $ P_{\rm N}$  is $ 0.9 \,  (\mathrm{mK})^2 (\MpcOh)^3  $ for MeerKAT versus  $ 30 \,  (\mathrm{mK})^2 (\MpcOh)^3  $ assumed in R10k2.  For $\xi_{r \parallel} $ and $\mu$-wedge, we find that the constraints in  MeerKAT are tighter than in R10k2 by different extents, while for the multipoles, the transverse BAO constraint is tighter but the radial BAO constraint is weaker relative to R10k2.  \citet{KennedyBull_2021} conducted a BAO forecast for MeerKAT using the multipole correlation functions. Their error bars on $\alpha_\perp$  and $ \alpha_\parallel $  are about 0.12 and 0.09, respectively. There are two main reasons for the discrepancy with ours. First, and more importantly, linear BAO is used in their modeling.  In fact, their constraints are similar to our R10k2 results shown in Table \ref{tab:multipole_parameter_l024} because the BAO damping can be neglected at such high redshift.   At low redshifts such as the MeerKAT redshift, the radial BAO feature is significantly damped (the transverse BAO is already damped by the beam), and this causes the radial BAO constraint to be weakened. Second, we include higher multipoles beyond the quadrupole.  %Third, their parameters used, such as $k_{\rm fg}$ and $ R_{\rm beam } $ are slightly different from ours. 

In \citet{Villaescusa-Navarro_etal2017}, the radial power spectrum $ P_\parallel $ was used to forecast the BAO constraint for SKA-mid. They derive  $(S/N)^2 $ of 9990, while our radial correlation function $\xi_{r \parallel}$ yields  $ (S/N)^2 = 27800 $. Our higher $(S/N)^2$ can be attributed to the fact that  $\xi_{r \parallel}$ has been optimized with respect to  $r_{\perp \rm max} $.

While our forecasts are still idealistic, e.g., the Gaussian beam approximation and the assumption that  there are no residual effects after foreground removal except the suppression of the modes by Eq.~\eqref{eq:Bfg_model}, nonetheless, our model has captured the essence of these two important observational effects. With the (minor) differences in the forecast implementations in mind, our results broadly agree with others' and seem to suggest that our optimized statistics yield more attractive constraints. Our work can serve as a useful guide to these experiments to further strengthen their constraining power.

\section{Conclusions}
\label{sec:conclusions}

There are a number of ongoing and forthcoming experiments aiming to measure the BAO using the 21 cm signals in the late universe. In this work, we study the methods for 21 cm IM BAO measurement in configuration space using mocks and theory, with the focus on the 21 cm IM observational effects caused by the beam and foreground removal.  We systematically compare the performance of three types of correlation functions for 21 cm IM BAO recovery, including the radial correlation function, multipole correlation function, and $\mu$-wedge correlation function.  Although these statistics are known in the literature, a quantitative comparison of their efficacy in BAO recovery was lacking, and it is the goal of this paper to fill this gap.

The telescope beam and foreground removal effects are unique to 21 cm IM observation and can induce strong anisotropies in the correlation function.  The beam smooths the correlation function in the transverse direction, and we model its form as Gaussian. The foreground removal effect is a consequence of 21 cm foreground cleaning, and it suppresses the long-wavelength modes in the radial direction.  In particular, these effects cause strong artifacts in the transverse and radial axes, respectively, and we discuss their formation using some simple analytic models in Sec.~\ref{sec:21cm_Pk}.

We develop a BAO fitting pipeline for each of these correlation function estimators. The theory templates are presented in Sec.~\ref{sec:correlation_functions}, and we derive their Gaussian covariance in Sec.~\ref{sec:covariances}.  To verify and calibrate our pipeline, we make use of the  \HI simulation from \citet{Avila_etal2022}. The mock catalogs use SAM simulations to model baryon and galaxy properties. The final \HI is obtained by post-processing the cold gas from the SAM simulation. From this 21 cm IM mock, we further add the beam and foreground removal effects.

We use the catalog to look for the optimal parameter choices for the three types of correlation functions under various observation conditions.  For the radial correlation function $ \xi_{r \parallel } $, we find that the BAO constraints tighten mildly with the transverse averaging scale, $ r_{\perp \rm max} $. The constraints saturate at large  $ r_{\perp \rm max} \sim 50 \MpcOh $, roughly independent of the $ R_{\rm beam}$ and $k_{\rm fg}$ values, and thus we adopt  $ r_{\perp \rm max} = 50 \MpcOh $ in defining  $ \xi_{r \parallel } $.  For the multipole correlation function, when the quadrupole is added to the monopole, the constraint tightens substantially. They are further strengthened, albeit with increasingly diminishing marginal return when the hexadecapole and hexacontatetrapole are also included.  We consider the combination of monopole, quadrupole, hexadecapole, and hexacontatetrapole in our analysis. The $\mu$-wedge correlation function allows the flexibility to cut  some problematic regions out. This is especially advantageous to the 21 cm IM correlation function as the strong artifacts caused by the observational effects are well localized in the regions close to the axes.  By removing both ends, i.e.,~$\mu$ close to 1 and near 0, although the error bars are enlarged, robust BAO constraints can be derived.  To achieve high S/N, we further divide the $\mu$-range into multiple wedges to enhance the resolution.  %We find that with these strategies, these statistics achieve robust and consistent constraints across various observation conditions.
\change{ We keep both the full-wedge (10 wedges) and trimmed-wedge (6 wedges) scenarios for comparison. These results are summarized in Fig.~\ref{fig:BAOfit_results}. }

We apply the statistics with the optimal settings to forecast the BAO constraints for the 21 cm IM experiments, including BINGO, MeerKAT, and SKA (Table \ref{tab:21cmexperiments_parameterconstraints} and Fig.~\ref{fig:all_telescopes_BAO_fit}).
\change{  We find that for the low-redshift experiments BINGO and MeerKAT, the $\mu$-wedge correlation function achieves the tightest constraint for both $ \alpha_\perp$  and  $\alpha_\parallel $. For SKA-mid, the radial correlation function delivers the tightest constraint on  $\alpha_\parallel$, while the multipole correlation function gives the smallest error bars on  $ \alpha_\perp$. Our results also indicate that the 10-wedge setting does not lead to more unstable results for these experiments, so it is preferred over the trimmed six-wedge one.  }  Although our analyses are still idealistic, our model captures the essence of the beam and foreground removal effects, and our results are expected to be a useful guide for these experiments.

%From the correspondence between the correlation function and power spectrum, we anticipate that the  $\mu$-wedge power spectrum to be more constraining than the other two statistics. In another vein, the analyses here can be easily ported to investigate the efficacy of different cross correlation function estimators on the BAO recovery. We leave the detailed analyses to future studies. 

\section*{Acknowledgments}
ZZ and KCC acknowledge the support by the National Science Foundation of China under the grant number 12273121 and 12533002, and the science research grant from the China Manned Space Project with CMS-CSST-2025-A02. 
SA has been funded by MCIN/AEI/10.13039/501100011033 and FSE+ (Europe) under project PID2024-156844NA-C22 and the RYC2022-037311-I fellowship.
BVG would like to acknowledge the support from the European Union (ERC StG, LSS BeyondAverage, 101075919) and the Comunidad de Madrid 2019-T1/TIC-12702 grant.

%% \end{thebibliography}
%\clearpage
\bibliographystyle{aasjournal}
\bibliography{reference}

\end{document}